\documentclass[journal]{IEEEtran}
\usepackage{algorithm}
\usepackage{algorithmic}
\usepackage{tikz}
\usepackage{xcolor}
\newcommand*\circled[1]{\tikz[baseline=(char.base)]{
            \node[shape=circle,fill,inner sep=1pt] (char) {\textcolor{white}{#1}};}}
 \usepackage{graphicx}
\usepackage{subfig}
\usepackage{float}
\usepackage{cite}

\begin{document}

\title{Zero Aware Configurable Data Encoding by Skipping Transfer for Error Resilient Applications}

\author{Chandan Kumar Jha,
        Shreyas Singh, Riddhi Thakker, Manu Awasthi,  and Joycee Mekie  
\thanks{This work was done when Riddhi Thakker was at the Indian Institute of Technology Gandhinagar.}
\thanks{Chandan Kumar Jha and Joycee Mekie are with the Discipline
of Electrical Engineering, Indian Institute of Technology Gandhinagar. e-mail: chandan.jha@iitgn.ac.in and joycee@iitgn.ac.in}
\thanks{Shreyas Singh is with the Discipline
of Computer Science and Engineering, Indian Institute of Technology Gandhinagar. e-mail: shreyas.singh@iitgn.ac.in}
\thanks{Riddhi Thakker was with the Discipline
of Information and Communication Technology from Dhirubhai Ambani Institute of Information and Communication Technology, Gandhinagar. She is now with Oracle India. e-mail: 201601124@daiict.ac.in}
\thanks{Manu Awasthi is with the Department
of Computer Science, Ashoka University, Sonepat. e-mail: manu.awasthi@ashoka.edu.in}
\thanks{© 2021 IEEE. Personal use of this material is permitted. Permission from
IEEE must be obtained for all other uses, in any current or future media,
including reprinting/republishing this material for advertising or promotional
purposes, creating new collective works, for resale or redistribution to servers
or lists, or reuse of any copyrighted component of this work in other works.}
}


\maketitle

\begin{abstract}
 Data transfer across DRAM channels accounts for nearly a quarter of the total energy consumption of DDR4 DRAMs. Modern applications 
with high bandwidth requirements further increase channel energy consumption.
However, channel energy consumption is dependent on data being transferred.
Pseudo Open Drain (POD) asymmetric termination, used in current DDR4 systems, consumes energy only when 1's are being transmitted over the channels. Many modern applications, including AI/ML ones are resilient to errors 
in data, and can work well with approximate data. This resilience can 
vary widely across and within applications, which provides a number 
of ways for exploiting these characteristics to save data transfer energy
across the DRAM channel. However, all DRAM data encoding schemes have been 
targeted towards applications that require exact data and are not approximation
resilient.

In this paper, we propose Zero Aware Configurable Data Encoding by Skipping Transfer (ZAC-DEST), a data
encoding scheme to reduce the energy consumption of DRAM channels, 
specifically targeted towards approximate computing and error resilient
applications. ZAC-DEST exploits the similarity between recent data transfers across
channels and information abut error resilience behaviour of applications to
reduce on-die termination and switching energy by reducing the number of 1's
transmitted over the channels. ZAC-DEST also provides a number of knobs for trading
off application's accuracy for energy savings, and vice versa, and 
can be applied to both training and inference.

We apply ZAC-DEST to five machine learning applications.
On average, across all applications and configurations, we observed a reduction of $40$\% in termination energy and $37$\% in switching energy as compared to the state of the art data encoding technique BD-Coder with an average output quality loss of $10$\%. We show that if both training and testing are done assuming the presence of ZAC-DEST, the output quality of the applications can be improved upto 9$\times$ as compared to when ZAC-DEST is only applied during testing leading to energy savings during training and inference with increased output quality.
\end{abstract}


\section{Introduction}
\label{sec:intro}
DRAMs are an integral component of memory systems~\cite{jacob2010memory,wang2014bigdatabench,young2018accord,li2018performance}. DRAM energy accounts for approximately $46$\% of the total system energy consumption~\cite{elmore2016analysis, barroso2009datacenter}.
The energy consumption of the DRAM I/O channel contributes $25$\% of the total DRAM energy due to off-chip communications~\cite{david2011memory,behnam2020stfl}. To reduce DRAM I/O energy consumption, asymmetric I/O termination mechanisms like Pseudo Open Drain (POD) and Low Voltage Swing Terminated Logic (LVSTL) have been implemented~\cite{jedec,jedecl,sohn20131,cho2013sub}. These mechanisms help in reducing energy consumption of DRAM channels~\cite{jedec,jedecl,sohn20131,cho2013sub}. This happens because asymmetric termination mechanisms dissipate energy for only one of the bits during data transfer, i.e. for bit-0 in LVSTL and bit-1 in POD~\cite{jedec,lee2018reducing,seol2016energy}. 

Error resilient applications in the domain of machine learning, object recognition, image/video processing etc. have opened up a plethora of
possibilities to optimize current
computing and memory systems~\cite{zeinali2017progressive,venkataramani2020efficient,venkataramani2015approximate}.
The error resilience of applications is exploited by introducing
approximation in computation or data to reduce energy and/or improve performance. As a result, the applications are able 
to achieve the same level of performance and accuracy, with sometimes 
significant amount of approximation introduced into the data, which 
enables us to explore trade-off between accuracy and energy savings. 
Previous research explored approximate data encoding for serial data transfer
in embedded systems~\cite{pagliari2017approximate, pagliari2016serial,kim2017axserbus}. Recent works have also explored approximate compression and decompression of
data~\cite{boyapati2017approx,stevens2018axba}. 

DRAM I/O energy consists of two components, namely termination and switching.
Termination energy is consumed in DRAM channels as a result of on-die
termination. Switching energy is consumed due to charging of DRAM channels during
data transfer. Termination energy in POD, used in DDR4 DRAMs, is directly
proportional to the number of $1$’s being sent over the DRAM channel. Bit value
$1$ is sent using $0$V and bit value $0$ is sent using
$V_{dd}$V~\cite{lee2018reducing}, where $V_{dd}$ is the supply voltage. The
number of 1’s in a data word, also called its hamming weight, has a positive
correlation with termination energy~\cite{jedec}. On the other hand, switching
energy is proportional to the number of 1 to 0 (charging) transitions. For 0 to 1
(discharging) transitions, no current is drawn from the supply
voltage~\cite{jedec,lee2018reducing,seol2016energy}. In most cases, reducing
hamming weight also leads to a reduction in switching count, thus reducing
switching energy as well~\cite{seol2016energy}. In modern DRAMs, which deploy one
of the two termination schemes, the termination energy has become the dominant
source of energy consumption in DRAM channels~\cite{jedec,seol2016energy}. Thus,
in recent years, research has focused on reducing termination energy by reducing
the number of 1’s sent across the channel~\cite{lee2018reducing,seol2016energy}.
However, to the best of our knowledge, there exists no prior research which 
looks into encoding the data approximately between DRAM channels and processors. This is the first work to exploit approximate computing to benefit while performing data transfers.

In this paper, we propose \textbf{ZAC-DEST} (\underline{Z}ero \underline{A}ware \underline{C}onfigurable \underline{D}ata
\underline{E}ncoding by \underline{S}kipping
\underline{T}ransfer), an energy efficient data
encoding scheme for DRAM channels for approximate computing
applications. DEST extends existing data encoding schemes
for exact data transfers - Bitwise Difference Encoding (BD-Coder)~\cite{seol2016energy} and 
Dynamic Bus Inversion (DBI)~\cite{stan1995bus} with additional optimizations options 
provided by error and approximation tolerant applications
to provide a further reduction in
termination and switching energies of DRAM I/O.

Transfer of approximate data across the DRAM channel leads to energy savings at the cost of output quality loss in applications. An error resilient application can tolerate varying amounts of approximation, i.e. there is a trade-off between output quality and energy consumption. ZAC-DEST introduces three tuning features -- i) \emph{Similarity Limit}, ii) \emph{Truncation}, and iii) \emph{Tolerance} which allows it to introduce a varying range of
approximations in data sent across the DRAM channel, and as a result, allows
for interesting trade-offs to be made by the architect or the application 
programmer. In most applications increasing the approximation leads to
an increase in  energy savings with a reduction in output
quality~\cite{simonyan2014very}.

Depending upon the
application, the acceptable output quality may vary. The
tuning features in ZAC-DEST allow it to be tailored for
obtaining acceptable output quality, while achieving significant energy contributions. Overall, in this paper,
we make the following contributions. 

\begin{itemize}
\item To  the  best  of  our  knowledge,  ZAC-DEST  is  the  first proposal  for  an encoding  mechanism  for  transferring data across DRAM channels geared specifically towards error resilient applications. ZAC-DEST extends the data encoding schemes designed for exact applications to provide and average savings of 40\% in  termination  and 37\% in  switching energy off-chip DRAM channels across five machine learning applications.

\item We  augment  the  existing  encoding  mechanisms  (BD-Coder) with two additional policies  that improves the BD-Coders’  table  update mechanisms.  In  addition, ZAC-DEST handles transfer of zeros across the channel separately,  which  is  useful  for  reducing  data  transfer energy when data to be transferred has majority zeros. On average the modified BD-Coder consumes 25\% lesser energy as compared to the original BD-Coder.

\item ZAC-DEST incorporates multiple knobs to trade off accuracy and DRAM channel transfer energy in error resilient applications. : Similarity Limit (exploits similarity between recent data transfers), Truncation(removing  bits  that  do  not  affect  output  quality), and Tolerance (masking  bits  that  cannot  tolerate approximation),  making  it an  ideal  candidate  for  use  in  approximate  processors. These knobs can be varied to obtain the desired accuracy and we have explored them in detail. We have also implemented ZAC-DEST design in UMC 65nm. The area overhead of ZAC-DEST  over BD-coder is 15\%.

\item We developed a framework that allows for the evaluation of DEST on error resilient machine learning applications. We also evaluated five different machine learning applications namely: i) ImageNet inferencing, ii) CIFAR-100 training and inferencing, iii) Eigenfaces, iv) Color quantization using K-Means and v) SVM. For each of the applications, we observe a reduction of  hamming energy by 39\%, 34\%, 44\%, 47\%, and 36\% respectively.

\item Finally, we demonstrate that inference accuracy of image classification of CIFAR-100 dataset using ResNet-110  can  increase  on an average 24\% (by  upto  9$\times$)  when  ZAC-DEST  is applied to  data  transfer  from  DRAM  during  both  training and inference phases, as opposed to application of ZAC-DEST to only the inference process data transfers. Thus, not  only  can  ZAC-DEST be  exploited  to  provide energy savings during both training and testing, but also improve the output accuracy.
\end{itemize} 

\section{Motivation}\label{sec:motivation}
\subsubsection*{\underline{Error Resilient Applications}}
\begin{figure}[]
\centering
\captionsetup{justification=centering}
\begin{tabular}{cccc}
\subfloat[PSNR=Inf]{\label{lena}\includegraphics[width=0.2\linewidth]{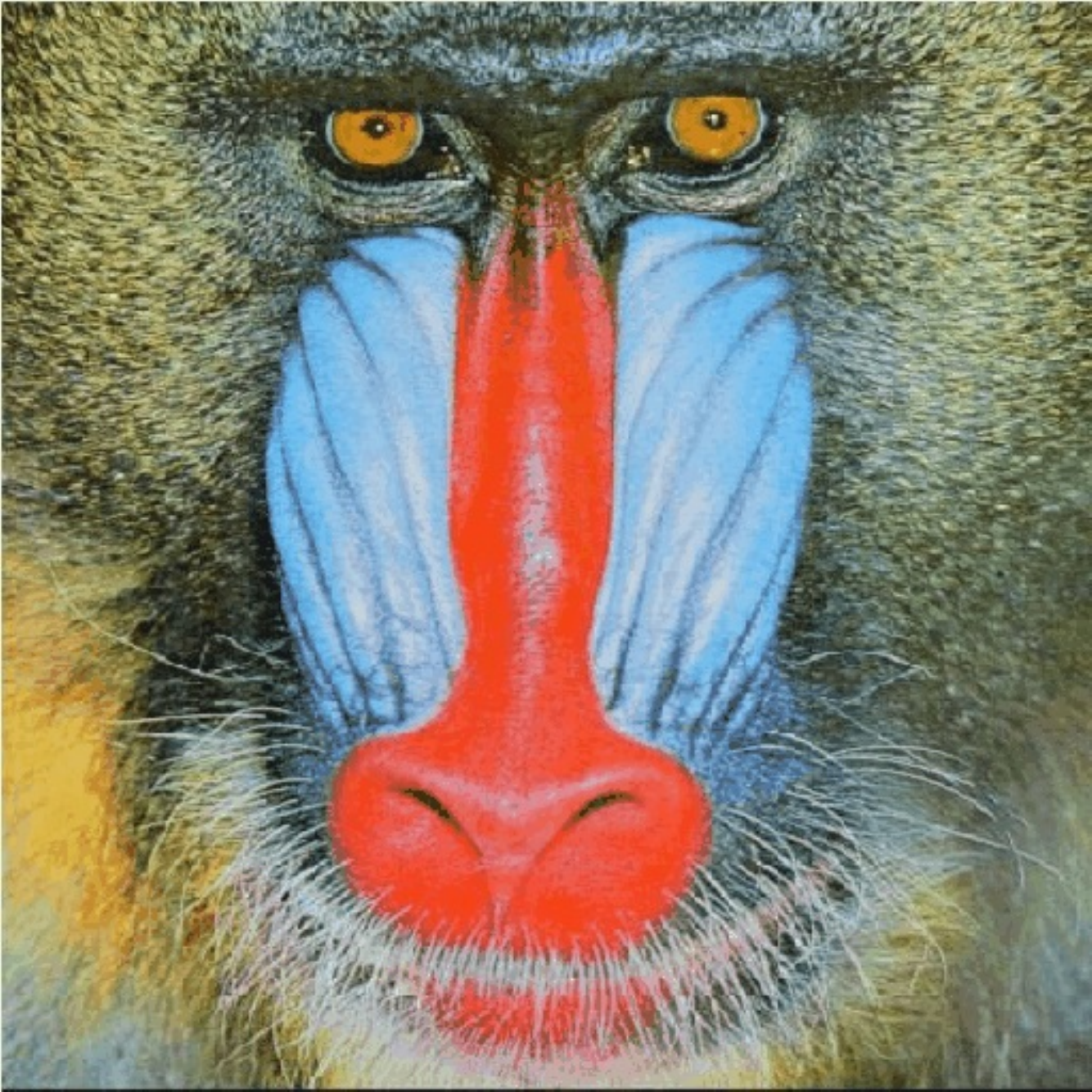}} &
\subfloat[PSNR=36]{\label{lenaflip}\includegraphics[width=0.2\linewidth]{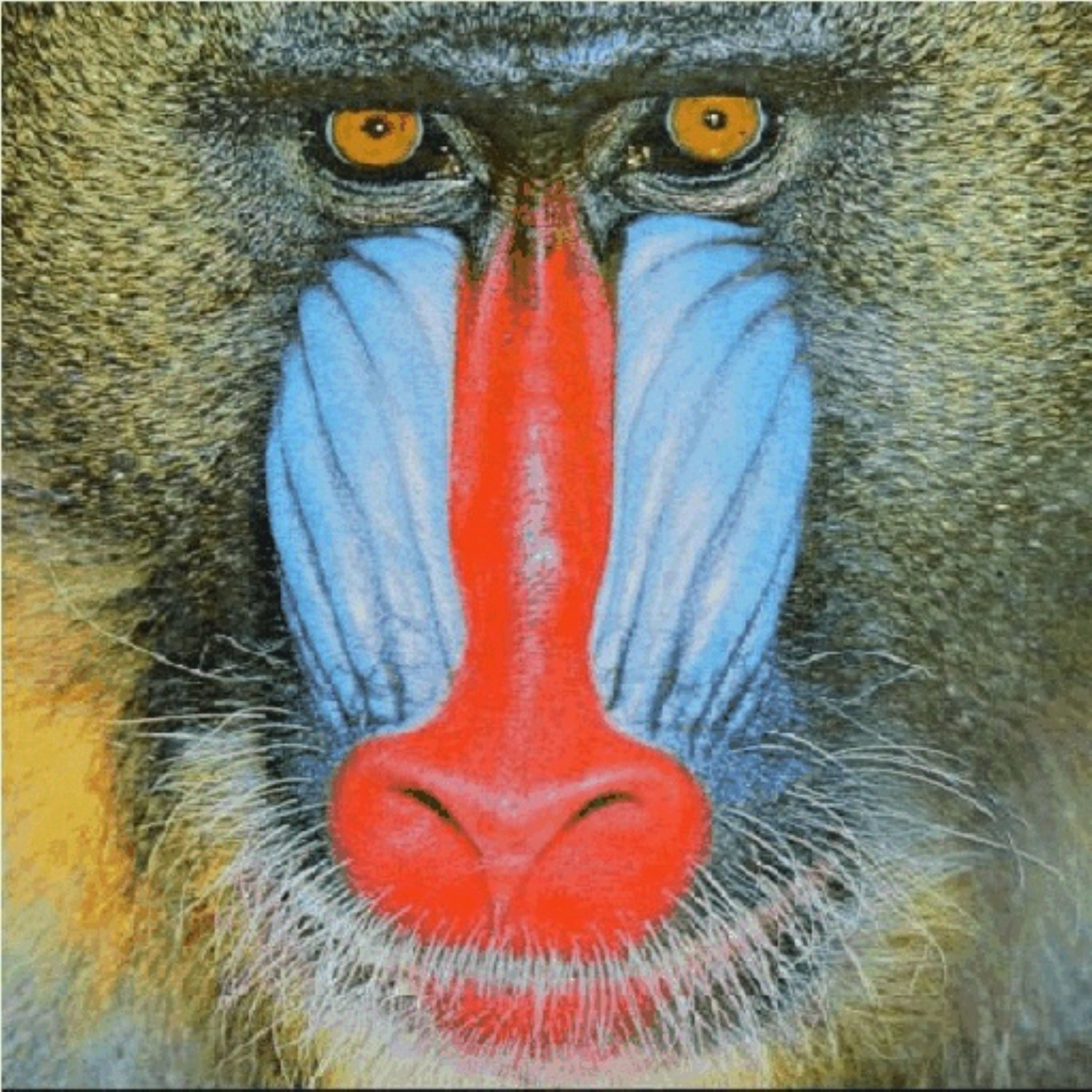}} &
\subfloat[PSNR=32]{\label{lenabadflip}\includegraphics[width=0.2\linewidth]{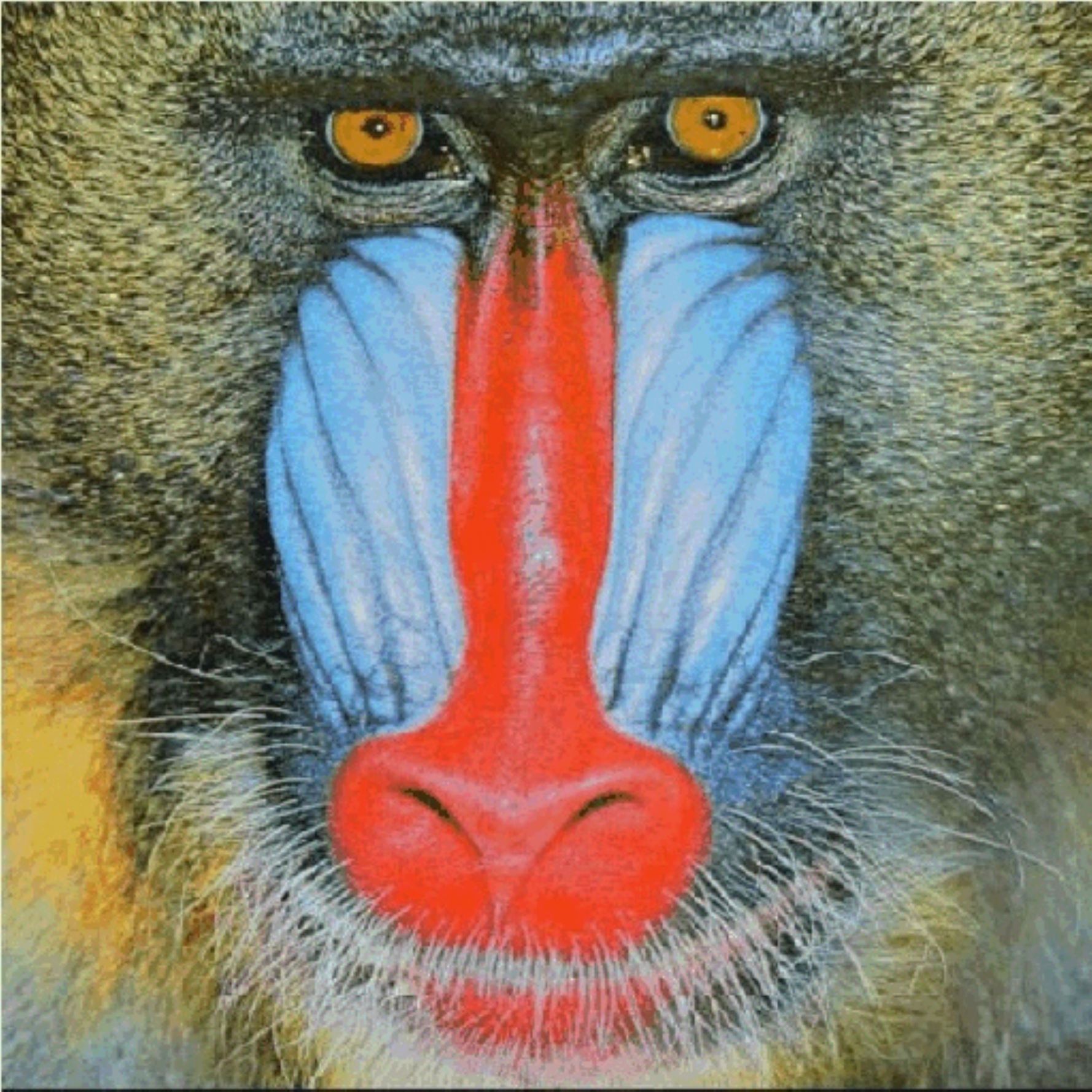}} \\
\end{tabular}
\caption {\protect\subref{lena} Original \protect\subref{lenaflip} 20\% 1's flipped in LSBs \protect\subref{lenabadflip} 40\% 1's flipped in LSBs}
\label{eval}
\end{figure}

Various recognition, mining, and synthesis applications are
resilient to some degree of approximation in data and computations~\cite{chippa2013analysis, esmaeilzadeh2012architecture, esmaeilzadeh2012neural,boyapati2017approx}. Machine learning
applications for object detection, image recognition,
etc. have also shown robustness towards
errors in data and computations~\cite{park2018blinkml,chen2018exploiting,cai2017deep,gupta2015deep,han2016eie,sze2017efficient,albericio2016cnvlutin,reagen2016minerva}. Thus, there exists a wide variety of applications where approximation can be traded off for energy savings. We
demonstrate error resilience in images with the help of an example image,
shown in Fig.~\ref{lena}. Every pixel of the
image is of an 8-bit entry. To introduce approximation in data, the 1's
in the last 4-bits of pixels were flipped to 0's. The percentage of
flipped 1's in Fig.~\ref{lenaflip} is 20\% and
Fig.~\ref{lenabadflip} is 40\%. Peak signal to
noise ratio (PSNR), a quality metric used to measure the similarity between images, is 36 for 
Fig.~\ref{lenaflip} and 32 for Fig.~\ref{lenabadflip},
higher PSNR is better~\cite{mittal2016survey}. For most images, PSNR $\geq$ 30 is acceptable~\cite{mittal2016survey} as it is indifferentiable to the human eye. This shows the error resilience of images towards bit flips. In later sections, we will show that when these kinds of approximated images are fed as input to error resilient applications, the output quality loss is minimal. Hence, there is an opportunity to reduce energy by approximating the data.

 \subsubsection*{\underline{Energy Consumption by DRAM I/O}}
\begin{figure}[]
\captionsetup{justification=centering}
\centering
\includegraphics[scale =0.25]{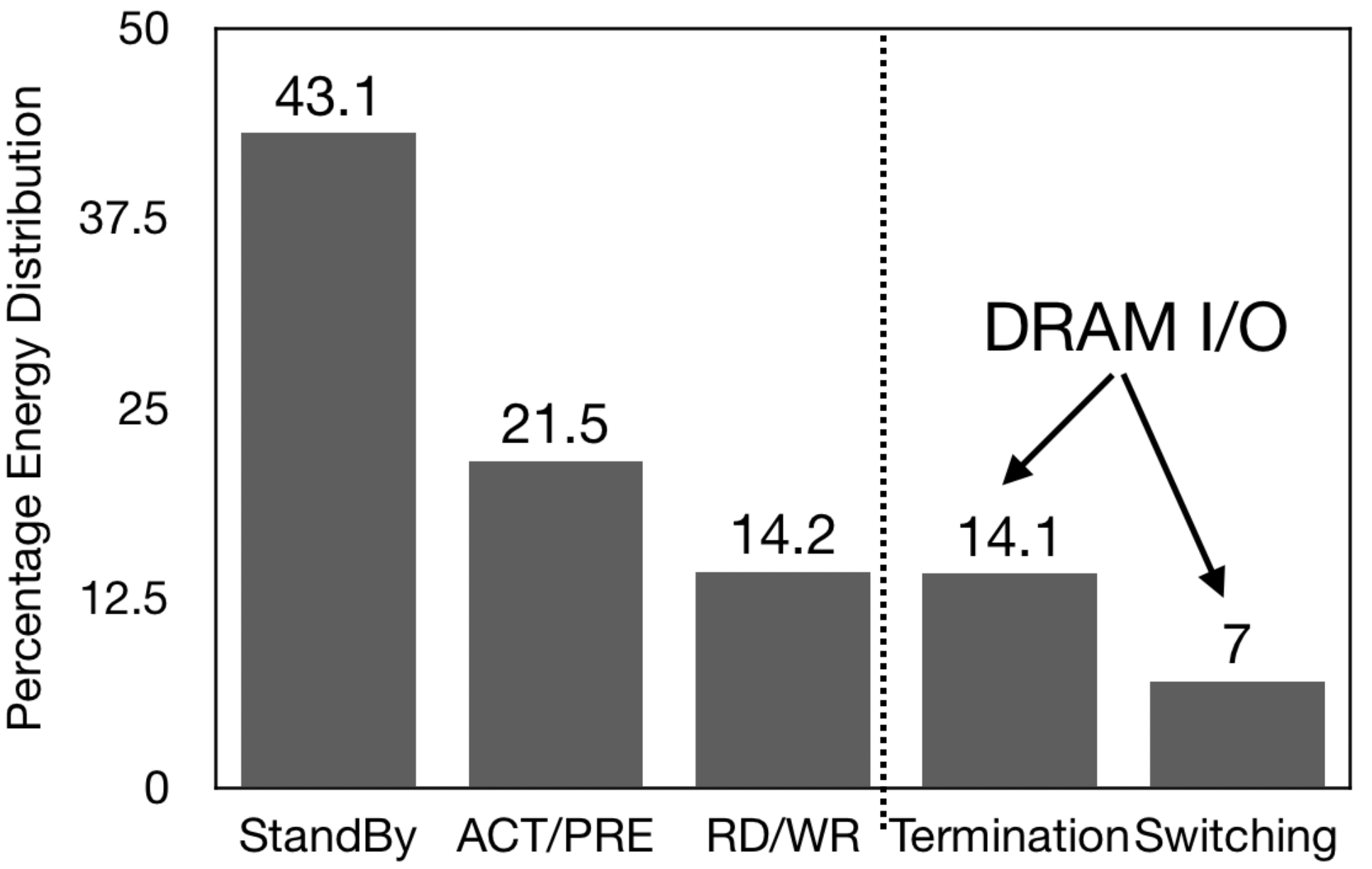}
\caption{Energy dissipation breakdown in DDR4 DRAM sub-system~\cite{seol2016energy}}
\label{ioenergy}
\end{figure}

A breakdown of energy consumption
of various DDR4 DRAM components was provided in~\cite{seol2016energy}, and is shown in Fig.~\ref{ioenergy}. We observe
that DRAM I/O energy (termination + switching) accounts
for 21\% of the total DRAM energy consumption. The termination
energy accounts for 67\% of the total DRAM I/O energy while the switching energy accounts for the rest. Prior
research has focused mostly on reducing termination energy~\cite{seol2016energy, lee2018reducing}. Furthermore, DRAM I/O energy is predicted to
worsen in the future as it is unaffected by scaling~\cite{seol2016energy,lee2018reducing}. This makes it crucial to devise techniques to reduce the DRAM I/O energy consumption.

\begin{figure}[]
\captionsetup{justification=centering}
\centering
\includegraphics[width=\linewidth]{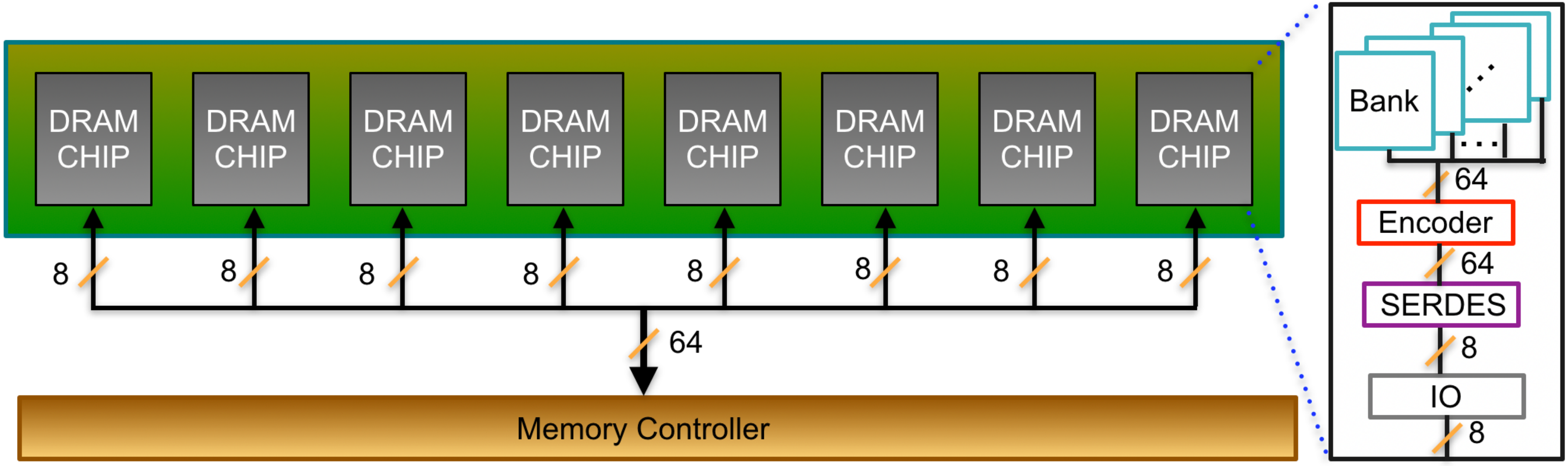}
\caption{Overall Data encoding-decoding structure}
\label{drammain}
\end{figure}
\section{Background and Prior Work}\label{sec:background}
\subsubsection*{\underline{DRAM Data Transfer}}
The data over the DRAM channel is transferred in 64 byte
(\emph{cache line}) granularity. DRAM data bus is 64-bit wide,
i.e. there are 64 physical lines from DRAM DIMM to the memory controller for the transfer of data.
There are other physical lines for the transfer of error correction codes, control commands, etc. The 64 byte cache line is transferred in 8 bursts of 64 bits each (\emph{assuming each chip is} x8)~\cite{ghose2018your}. For a 64-bit burst, the overall structure for data
encoding is shown in Fig.~\ref{drammain}. The encoder
is situated between the DRAM chip and the I/O bus, while the
decoder is located between the I/O bus and the memory
controller. In DRAMs, while transferring the bit value 
0, the DRAM channel is connected to $V_{dd}$ and for bit value 1, connected to $GND$~\cite{lee2018reducing}. 
\begin{figure}[]
\captionsetup{justification=centering}
\centering
\includegraphics[width=\linewidth]{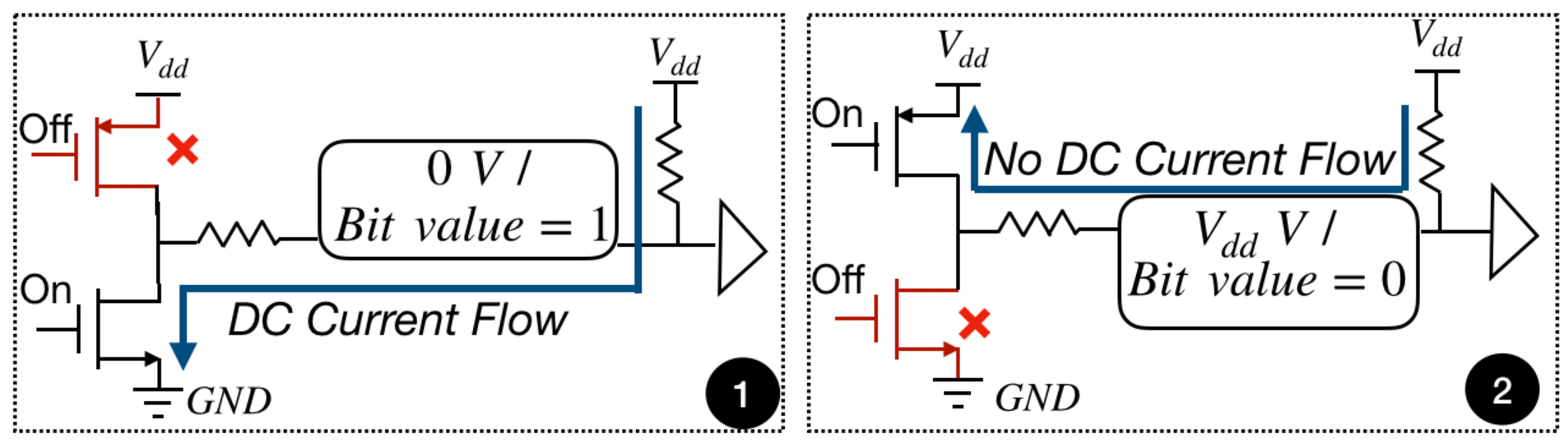}
\caption{Pseudo Open Drain I/O Termination}
\label{pod}
\end{figure}
\subsubsection*{\underline{DRAM I/O
Termination and Switching Energy}} POD I/O termination is a widely used termination
scheme in DDR4 DRAMs~\cite{jedec}. The termination energy is a
result of the POD I/O scheme. Due to asymmetric
design, it consumes a significant amount of energy, which depends
on termination resistance,  while transferring a bit value
1~\cite{lee2018reducing,seol2016energy}. This is due to the direct path between $V_{dd}$ and $GND$ as shown in
Fig.~\ref{pod}~(\circled{1}). This current accounts for the
termination energy in DRAM I/O's. When bit value 0, is
transmitted there is no current flow as shown
in Fig.~\ref{pod}~(\circled{2}). 
Transferring bit value 1 can consume 13.75~mA additional
current as compared to bit value 0~\cite{jedec}. Thus termination energy is directly proportional to the number of 1's transferred over the DRAM channel. Switching energy is proportional to the number of 1 to 0 (charging) transitions. The energy consumption as a result of switching is obtained using $E=CV_{dd}^2$, where $C$ is the capacitance and $V_{dd}$ is the supply voltage. The typical value of $C$ per channel is $15pF$~\cite{seol2016energy}. 
\subsubsection*{\underline{Bitwise Difference Coder (\emph{BD-Coder})}}\label{bdc}
\begin{figure}[]
\centering
\captionsetup{justification=centering}
\begin{tabular}{cccc}
\subfloat[]{\label{sendbd}\includegraphics[ width=0.48\linewidth]{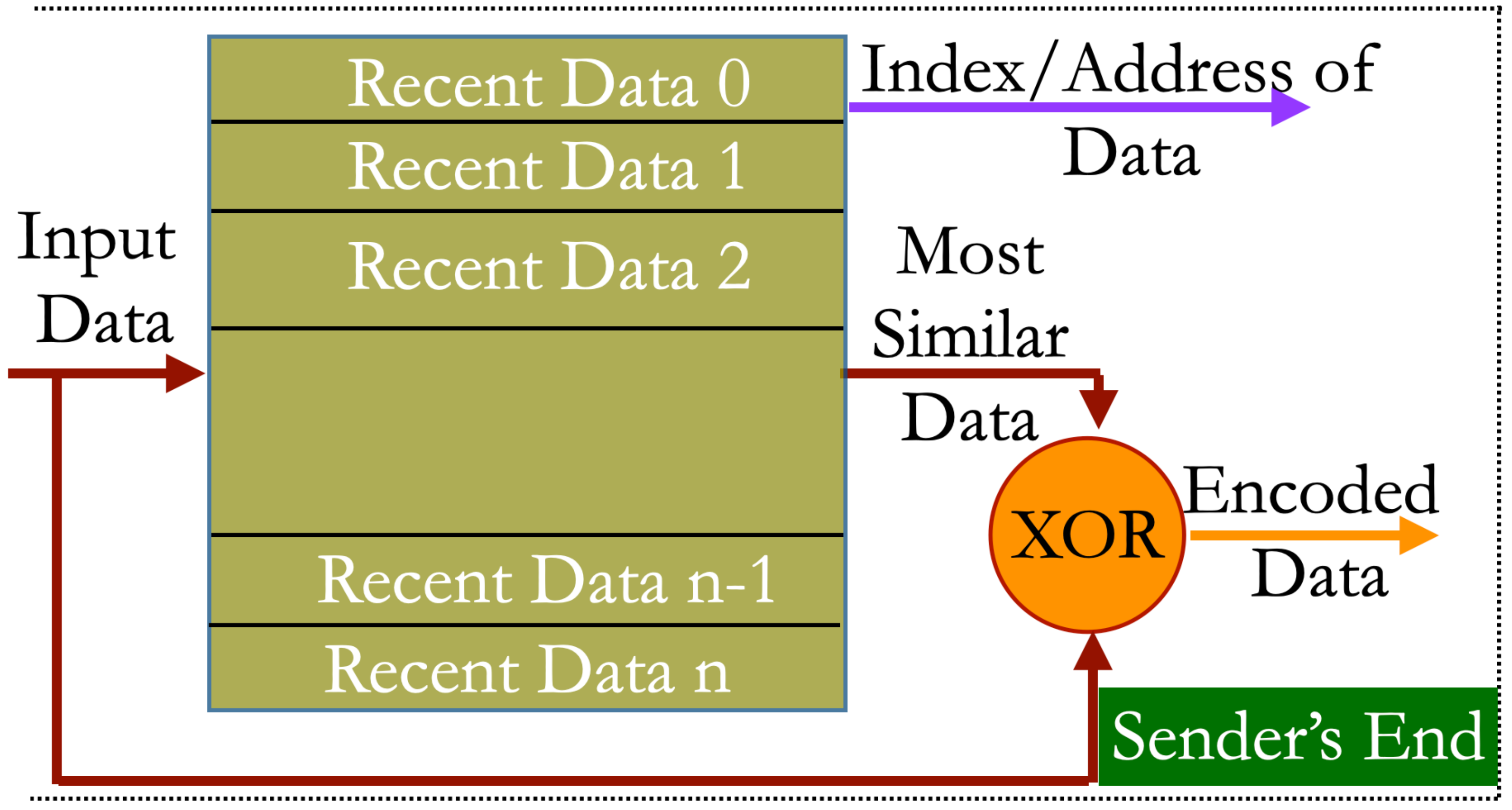}} &
\subfloat[]{\label{receivebd}\includegraphics[ width=0.5\linewidth]{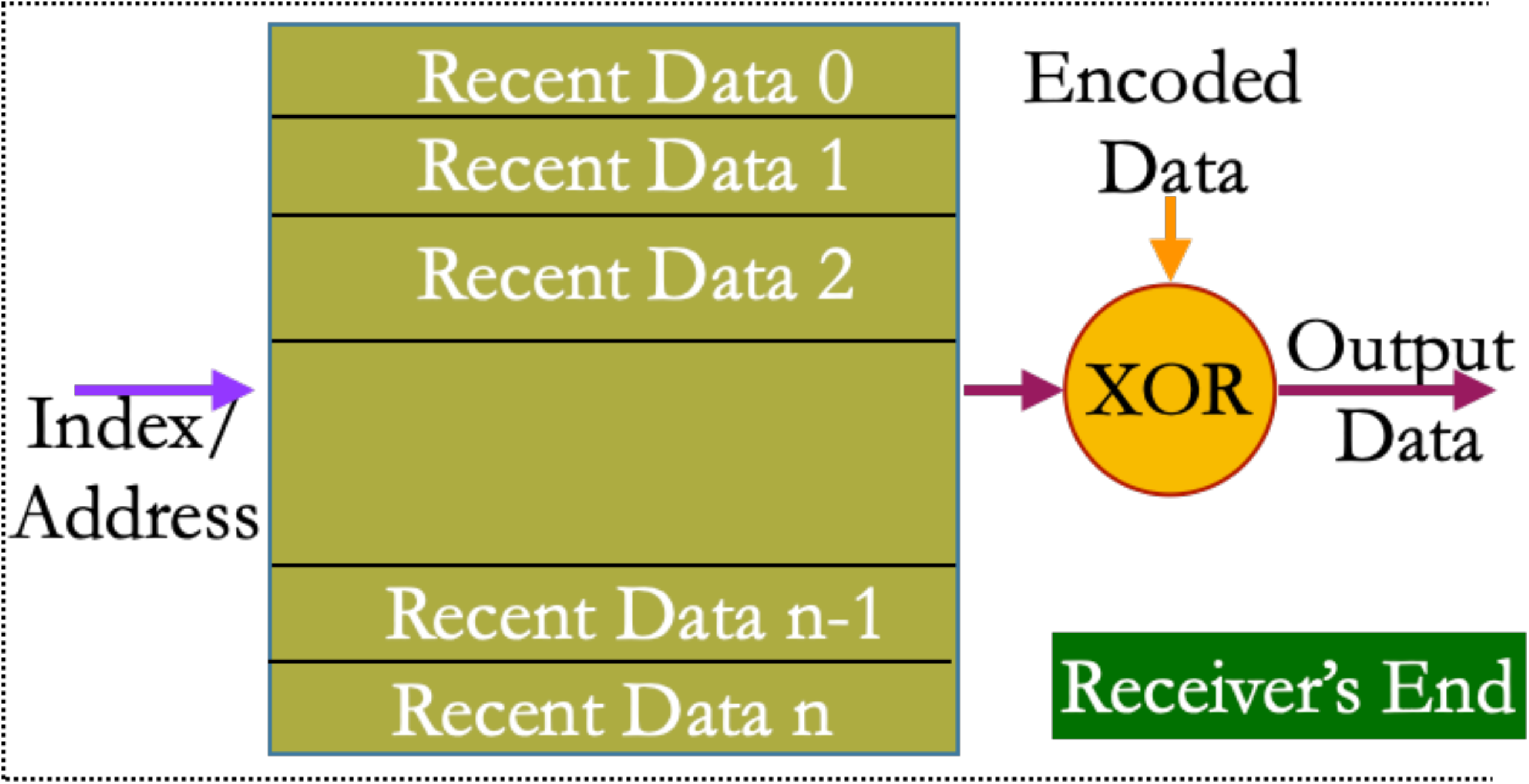}} &
\end{tabular}
\caption {BD-Coder~\protect\subref{sendbd}~Encoder~\protect\subref{receivebd}~Decoder}
\label{srbd}
\end{figure}

BD-Coder~\cite{seol2016energy} exploits data similarity between
recent data transfers to reduce DRAM I/O energy consumption. It
maintains a table (\emph{data table}) of recent data transfers at sender's (\emph{DRAM}) as well as receiver’s
(\emph{memory controller}) end. The data to be sent is first compared to all
the entries in the data table to find the most similar entry. To find the most similar entry, the
data to be sent is bitwise XORed with all the data table entries (\emph{XORing the same numbers gives a 0}) to reduce the
number of 1's. This new number of 1's (\emph{hamming weight})
in the XORed output is now compared to that of the original
data. If the XORed output has a smaller hamming weight, the
address/index of the most similar data (\emph{using a
separate line per chip}), along with the XORed output is
transferred over data lines. Otherwise, the original data
is sent over the data lines and the index lines send the address. If encoded data is received
at the receiver's end, it is XORed with the
data table entry pointed by the received address. Otherwise, the
original data is passed to the memory controller and the data table is updated with this data at both the sender and the receiver's end. The overall structure of the encoder and decoder in BD-Coder is shown in Fig.~\ref{sendbd} and Fig.~\ref{receivebd}, respectively.

\subsubsection*{\underline{Dynamic Bus Inversion (\emph{DBI})}}
DBI is widely used in DDR4 systems to reduce energy
consumption of the DRAM I/O~\cite{stan1995bus}. It is
applied at a granularity of $8$-bits. If more than $4$-bits out of $8$-bits
are 1's, DBI inverts the data being transferred. An
additional line is added per chip, i.e. 8 lines total, to convey if DBI has been applied. Thus, the
transmitted data always has at most four 1's leading to a
reduction in termination energy.

In the next section we will discuss ZAC-DEST, our proposed encoding scheme.

\section{\underline{Z}ero \underline{A}ware \underline{C}onfigurable \underline{D}ata \underline{E}ncoding by \underline{S}kipping \underline{T}ransfer (\emph{ZAC-DEST})}\label{sec4}

Most data encoding schemes try to reduce the DRAM I/O energy during data transfer. In this section, the mechanisms to exploit the error resilience when transferring data over the DRAM
channel to reduce DRAM I/O energy are explained. Since
termination energy is proportional to the 1's being transmitted, ZAC-DEST focuses reducing the number of 1's.
ZAC-DEST is built on top of current state of the art data encoding schemes for data transfer: a) BD-Coder~\cite{seol2016energy} and b)~DBI~\cite{stan1995bus}, which allows ZAC-DEST to be easily integrated with any encoding schemes built using similar design principles. 
\subsection{Leveraging Error Resilience} 
The similarity between recent data accesses remains the same irrespective of whether the application is
error resilient to input data. 
We leverage the error resilience of applications by introducing approximations with the goal of reducing
the number of 1's. The naive approach to introduce approximations will be to change all 1's to 0's when a request for approximate data is received. 
The key goal is to reduce the number of 1's being sent over the channel keeping the degree of approximation under check.

The amount of approximation that can be tolerated not only
varies widely across applications but also within
the same application. Thus, we need to provide a variety of
configurations that control the degree of approximation
introduced in data. Note that data transfers pertaining to instructions are \emph{never} approximated. Also, among the data, only the accesses that are known to be error resilient a priori are approximated. The information related to approximation, can be transferred over the
already existing address lines of the DRAM while transferring the
column address as column address have lesser bits as compared to the row address and it leaves some address lines unutilized~\cite{seol2016energy}. 

ZAC-DEST uses the same data table as in BD-Coder, shown in
Fig.~\ref{srbd}. Each data table, one per chip, holds `n' recent entries of 64-bits
transferred over the DRAM channel~\cite{seol2016energy}. We assume that the degree of approximation that can be tolerated by an
application will be known a priori and can be encoded in the
applications. The output quality for each workload will be defined in Section~\ref{sec:workloads}. The data to be sent is compared with all the entries in the data table. So,
if an application can tolerate a 25\% approximation in
data, 16 out of 64-bits can be approximated. The data to be
sent is compared to all the entries in the table to find the most
similar entry. The most similar entry is now checked to see if it differs from the original data by not more than 16-bits. If true, in place of actual data, all 0's are sent over the DRAM channel along with the index of the most similar entry, which is already present at the receiver's end. Note that this is
the same as best case scenario since we are not
transmitting any 1's. The only overhead is sending the
index of the receiver's data table at which the most similar entry is
stored. Here, the assumption is that number of 1's in
the index is very small as compared to the data. If
it would have been the case that the most similar entry has
less than 48 similar bits, we would have applied BD-Coder
on it i.e, the data would be sent without approximation. Thus, this encoding scheme fits very well on top of the existing data encoding scheme. 
 BD-Coder updates data table after every
transfer, which can lead to multiple entries having the same
value. In ZAC-DEST, we update the data table only when the \emph{exact} data
is transferred. This ensures no duplicate entries are present in the table. Since there are no duplicate data entries in ZAC-DEST, the probability of finding a most similar entry is higher, leading to further energy savings.
\subsection{Using the Unused } 
In frequent value (FV) encoding, the frequent values are encoded and sent as a one-hot encoded address and was targeted towards reducing switching energy~\cite{yang2004frequent}. ZAC-DEST differs from FV encoding as we have a separate encoding scheme and target termination energy. We will show how we exploit one hot encoding to further reduce the termination energy for ZAC-DEST. ZAC-DEST allows us to skip data transfer when a similar entry is
found in the data table. The only hiccup now is of transferring
the index (location) of the most similar entry. In BD-coder, a
separate line was used to transfer the index to the receiver. 
However, when ZAC-DEST is true, the skipping of data transfer during ZAC-DEST leaves the data
lines unused. These lines are used to our benefit for
sending the index of the entry in the {one-hot} encoded (OHE) format. For
example, in the worst case scenario which occurs when
transferring the index value $111111$ (i.e. $63$ in decimal) causes
six 1's to be sent. If the same is encoded in $64$-bit OHE, the
index sent will be `$0x8000000000000000$'. This reduces the number
of 1's down from six to one. Also, no additional lines are required since existing lines for data transfer are used to send the OHE index.

\section{ZAC-DEST Optimizations}\label{sec:advancements}
In this section, we discuss a separate addressing technique for zeros. We also discuss the support provided by ZAC-DEST for allowing configurability in  approximation within each data transfer. 
\subsection{Handling All Zeros}
Without any data encoding scheme, the transfer of 0's consumes the least amount of energy~\cite{lee2018reducing}. Hence, we must ensure that there are no overheads while transferring 0's.  
Thus, whenever a 64-bit data containing all 0's needs to be transferred, neither ZAC-DEST nor BD-coder applied to it. Also, unlike BD-Coder, which would update the table after every data transfer we do not add an entry in the data table when 0's are transferred which allows us to store unique data in the data table.
\subsection{Configurability in ZAC-DEST}
\subsubsection*{\underline{Similarity Limit}}
\emph{Similarity Limit}, as the name suggests refers to the number of bits that needs to same, between the data to be sent and the most similar entry, for ZAC-DEST to be true. We have included 4 different similarity limits in ZAC-DEST for evaluation purposes. These are 7, 13, 16, and 20 out of 64 bits which corresponding to 90\%, 80\%, 75\%, and 70\% similarity limit respectively. ZAC-DEST can be tuned to use any of the similarity limit values required by the application.
\subsubsection*{\underline{Tolerance}}
\emph{Tolerance} refers to the bits which \emph{cannot} be approximated. Even though we are proposing an encoding scheme for approximate applications, it may happen that approximating most significant bits (MSBs) may cause large errors in applications. These bits need to be transferred without approximation, irrespective of the similarity limit. Thus, the number of bits that can tolerate errors, in this case, will reduce. For example, if the data is 64-bits and a tolerance of 16 is required, the most significant 16 bits of data cannot be approximated. This will put a tighter constraint on  approximation and ZAC-DEST will be applied a fewer number of times. Support for a wide range of values that can be selected to tune required tolerance depending upon the data width is provided in ZAC-DEST. It is important to note that while having higher tolerance reduces energy savings, it does increase the output quality of the application. 
\subsubsection*{\underline{Truncation}}
Truncation refers to the removal of a fixed number of bits from the original data. In approximate computing, one of the most widely used approximation methodology is the removal of least significant bits (LSBs). Thus, it is useless to transfer these bits over the DRAM channel.  For example, if we have an 8-bit data of the form $01101111$ and we had to truncate 4 LSBs the data would change to $01100000$. 
In ZAC-DEST we incorporate truncation in the following way. If we have a 64-bit data and a truncation of 16 bits is required, the least significant 16 bits of the data will be ignored while finding the most similar entry. These bits will be replaced by 0's hereafter. The rest of the steps remain the same as that of ZAC-DEST. The overall algorithm for BD-Coder and ZAC-DEST is shown in Algorithm~\ref{alg_bde} and Algorithm~\ref{alg_dest}. 
\begin{algorithm}
\caption{BD-Coder Algorithm}
\begin{algorithmic} 
\STATE \underline{Definitions}
\STATE DCD- DRAM Chip Data
\STATE DS - Data sent over DRAM channels
\STATE DR - Data reconstructed at receiver end 
\STATE MSE - Most similar entry
\STATE BD-Coder- BDE
\FORALL{chip} 
\STATE Find MSE w.r.t DCD 
\STATE  \underline{Check for BDE}
\STATE  \textit{Condition for BDE to be True:}
\STATE  Hamm(DCD) $>$ Hamming Count of (MSE \emph{XOR} DCD) 
\IF {BDE condition TRUE}
\STATE DS : (MSE xored DCD) and Index of MSE
\STATE DR: DCD
\ELSE
\STATE DS : DCD
\STATE DR: DCD
\STATE Table Updated with DCD
\ENDIF
\ENDFOR
\end{algorithmic}
\label{alg_bde}
\end{algorithm}

\begin{algorithm}
\caption{ZAC-DEST Algorithm}
\begin{algorithmic} 
\STATE \underline{Definitions}
\STATE DCD- DRAM Chip Data
\STATE DCDT- DRAM Chip Data after Truncation
\STATE DS - Data sent over DRAM channels
\STATE DR - Data reconstructed at receiver end 
\STATE MSE - Most similar entry
\STATE MSET - Truncated most similar entry
\STATE ZAC-DEST -  Zero Aware Configurable Data Encoding by Skipping Transfer
\STATE MBDC - Modified Bitwise Difference Coder
\STATE DBI - Dynamic Bus Inversion
\FORALL{chip} 
\STATE Find MSE w.r.t DCDT \\
\COMMENT{\textit{Truncated bits are not used for comparison}}
\STATE  \underline{Check for Zeros}
\IF{DCDT == $0$}
\STATE return $0$ 
\ENDIF
\STATE  \underline{Check for ZAC-DEST}
\STATE  \textit{Condition for ZAC-DEST to be True:}
\STATE  Hamming Count of (MSET \emph{XOR} DCDT) $<$ Threshold and Tolerance bits are same
\IF{ZAC-DEST condition TRUE}
\STATE DS : OHE index of MSE 
\STATE DR:  MSET
\ELSE 
\STATE  \underline{Check for MBDC}
\STATE  \textit{Condition for MBDC to be True:}
\STATE  Hamm(DCDT) $>$ Hamming Count of (MSET \emph{XOR} DCDT) added to Hamming count of Index
\IF {MBDC condition TRUE}
\STATE DS : DBI (MSET xored DCDT) and Index of MSE
\STATE DR: DCDT
\ELSE
\STATE DS : DBI (DCDT)
\STATE DR: DCDT
\ENDIF
\STATE Table Updated with DCDT
\ENDIF
\ENDFOR
\end{algorithmic}
\label{alg_dest}
\end{algorithm}

\begin{figure*}[]
\centering
\captionsetup{justification=centering}
\begin{tabular}{ccccccc}
\subfloat[]{\label{camo}\includegraphics[height = 4.5cm, width =3.2cm]{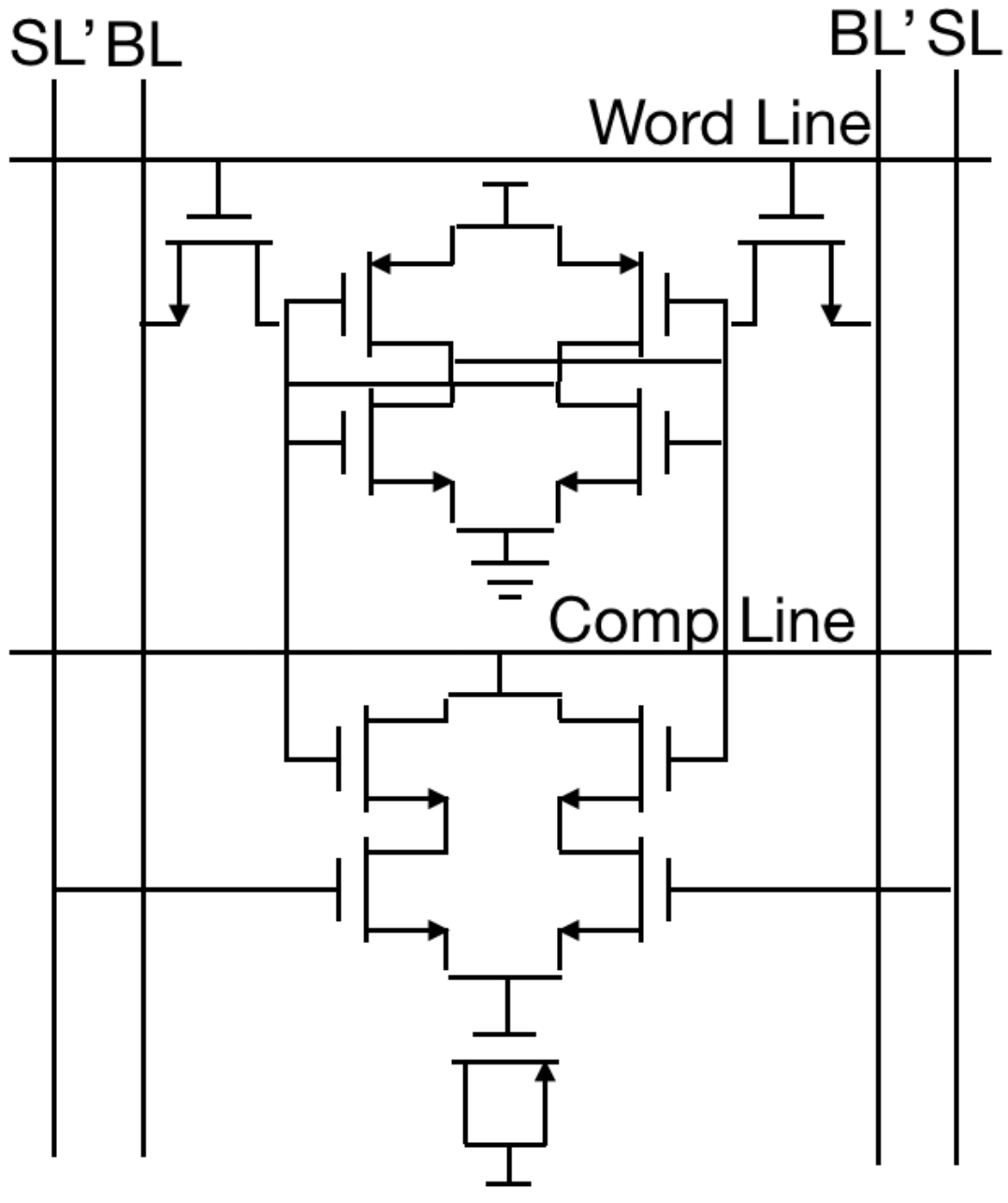}} &
\subfloat[]{\label{camp}\includegraphics[height = 4.5cm, width =3.2cm]{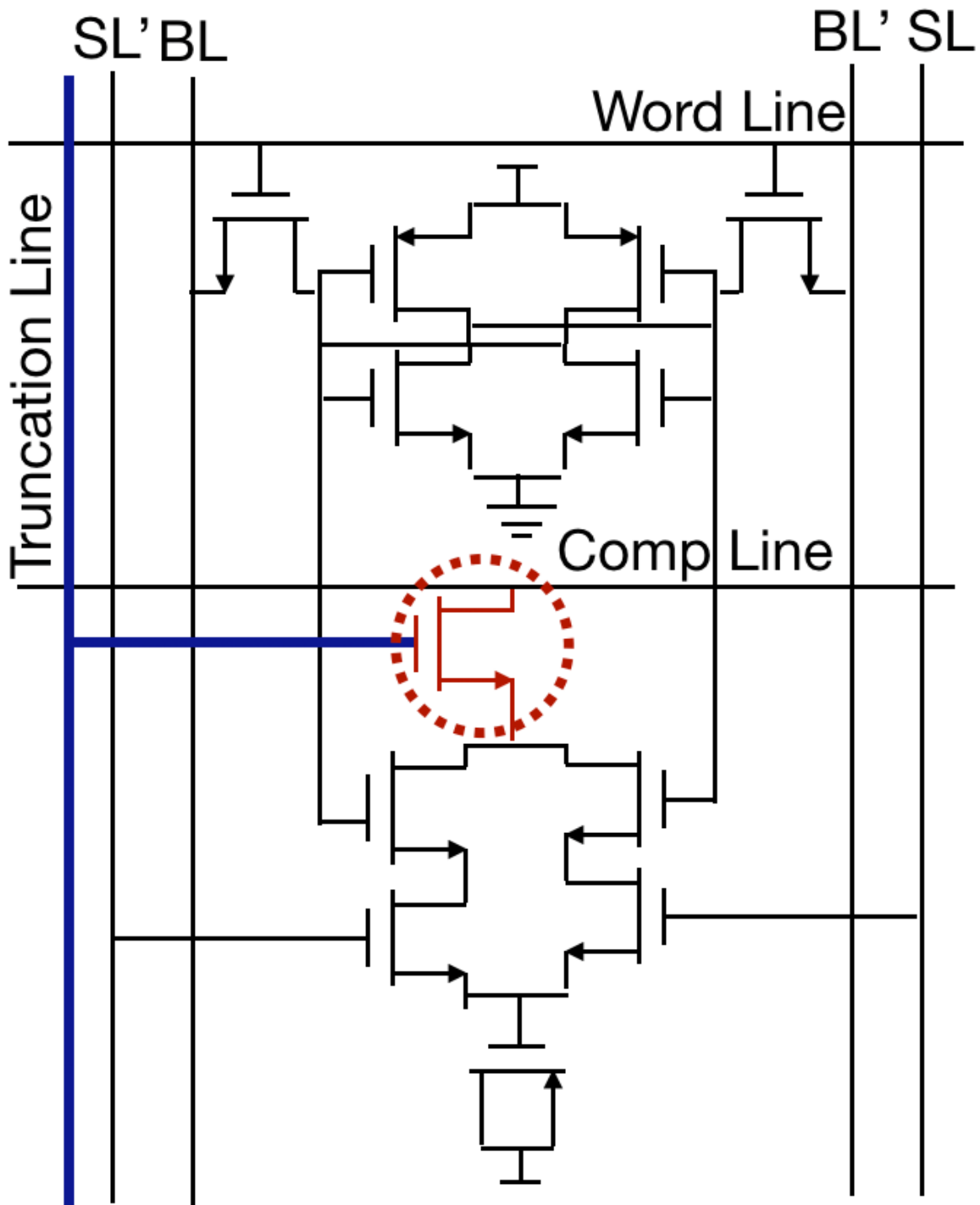}} &
\subfloat[]{\label{hamline}\includegraphics[height = 4.5cm, width =3.2cm]{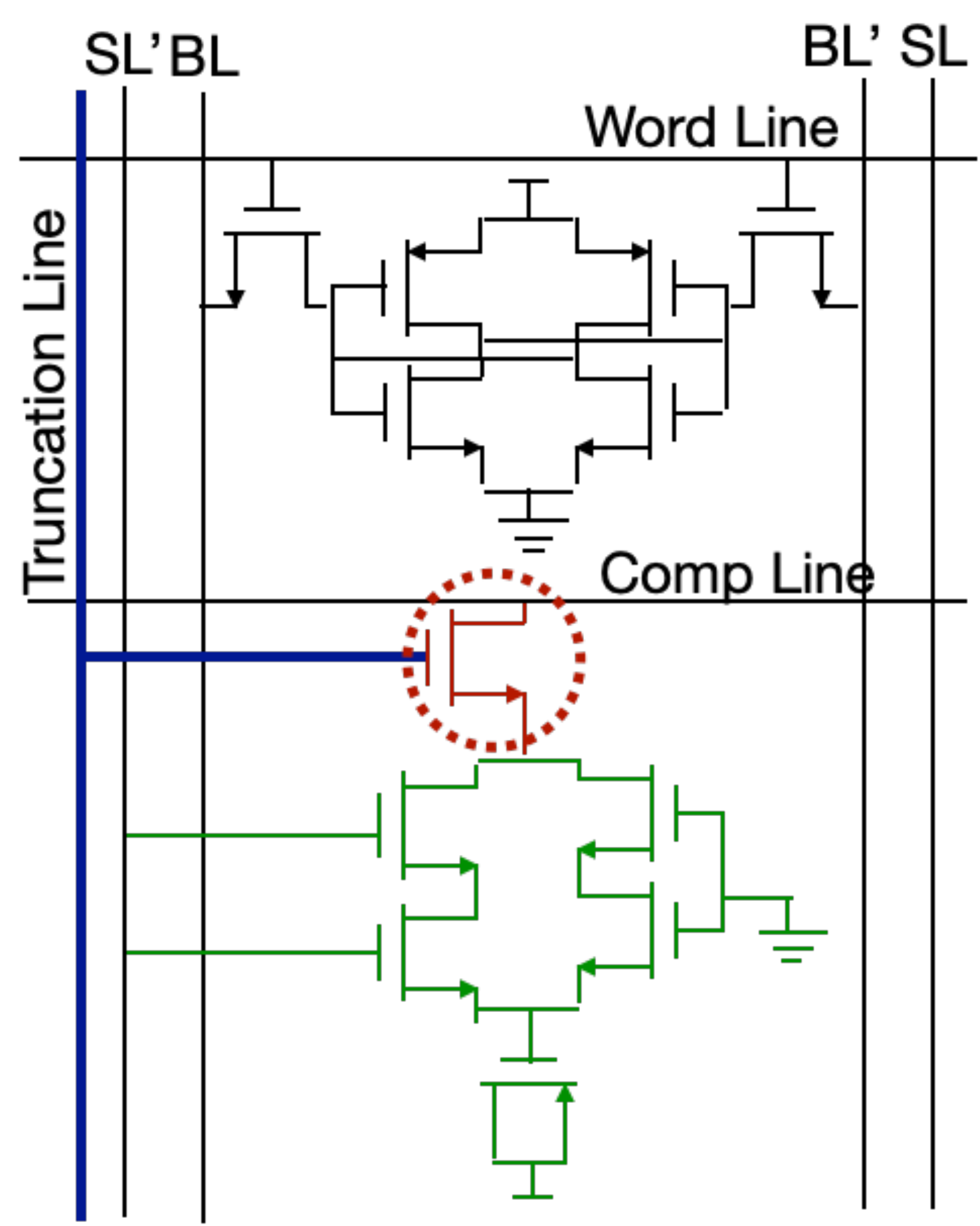}} &
\subfloat[]{\label{cbdcoder}\includegraphics[height =4.5cm, width =4.2cm]{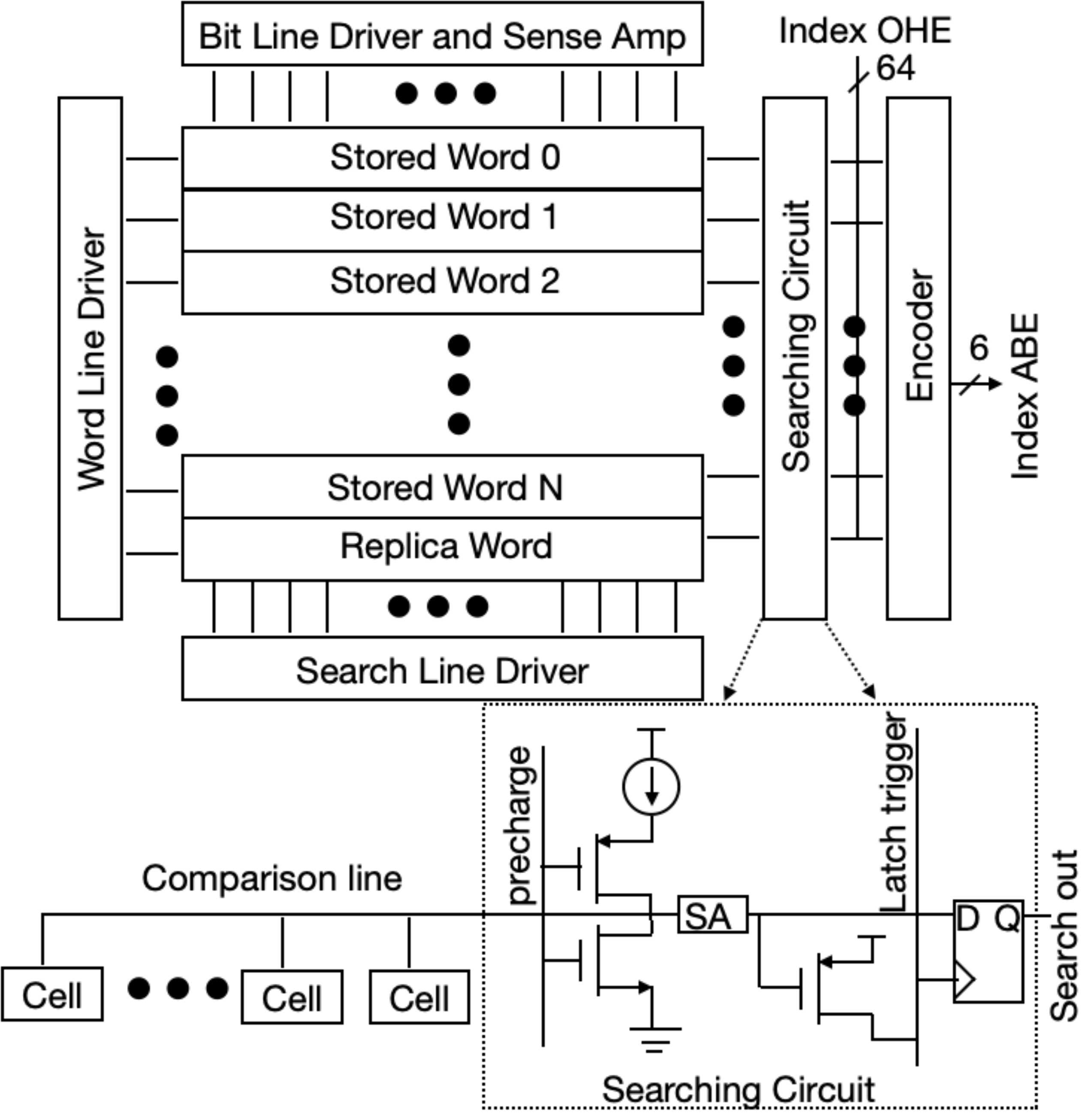}} &
\end{tabular}
\caption {\protect\subref{camo} Original CAM cell~\cite{seol2016energy} \protect\subref{camp} Modified CAM cell \protect\subref{hamline} Replica Word CAM Cell \protect\subref{cbdcoder} Modified BD Coder (MBDC) Data Table for ZAC-DEST }
\label{eval}
\end{figure*}
\section {ZAC-DEST Circuit Implementation}\label{sec:circuits}
The detailed circuit implementation of ZAC-DEST will be shown in this section. 
\subsubsection*{\underline{ZAC-DEST Data Table}}
We start with modifying the BD-Coder design. Fig.~\ref{camo} shows the NOR based binary content addressable memory (CAM) used to implement the data table. The data table in BD-Coder does the following \textbf{i)} stores recent data transfers, and \textbf{ii)} finds the most similar entry (MSE). \textbf{(i)} A 6-transistor based SRAM is used for storing the data in the CAM cell as shown in Fig.~\ref{camo}. This allows reading and writing data into the data table using BL and BL'. For \textbf{(ii)} a 5-transistor comparator is used and search is performed using SL and SL'.  The most similar entry is obtained using the comparator. For ZAC-DEST we have added one more feature \textbf{(iii)} for supporting truncation. For \textbf{(iii)} we add 1-transistor and an additional line, \emph{truncation line} in the CAM module as shown in Fig.~\ref{camp}. When the truncation line goes to 0, the NMOS connected to the line turns off and disconnects the comparator from the comparison line. This bit value connected to this line will then not be used for comparison. 
\begin{figure*}[]
\centering
\captionsetup{justification=centering}
\begin{tabular}{ccccccc}
\subfloat[]{\label{csub}\includegraphics[height = 3.5cm, width =7cm]{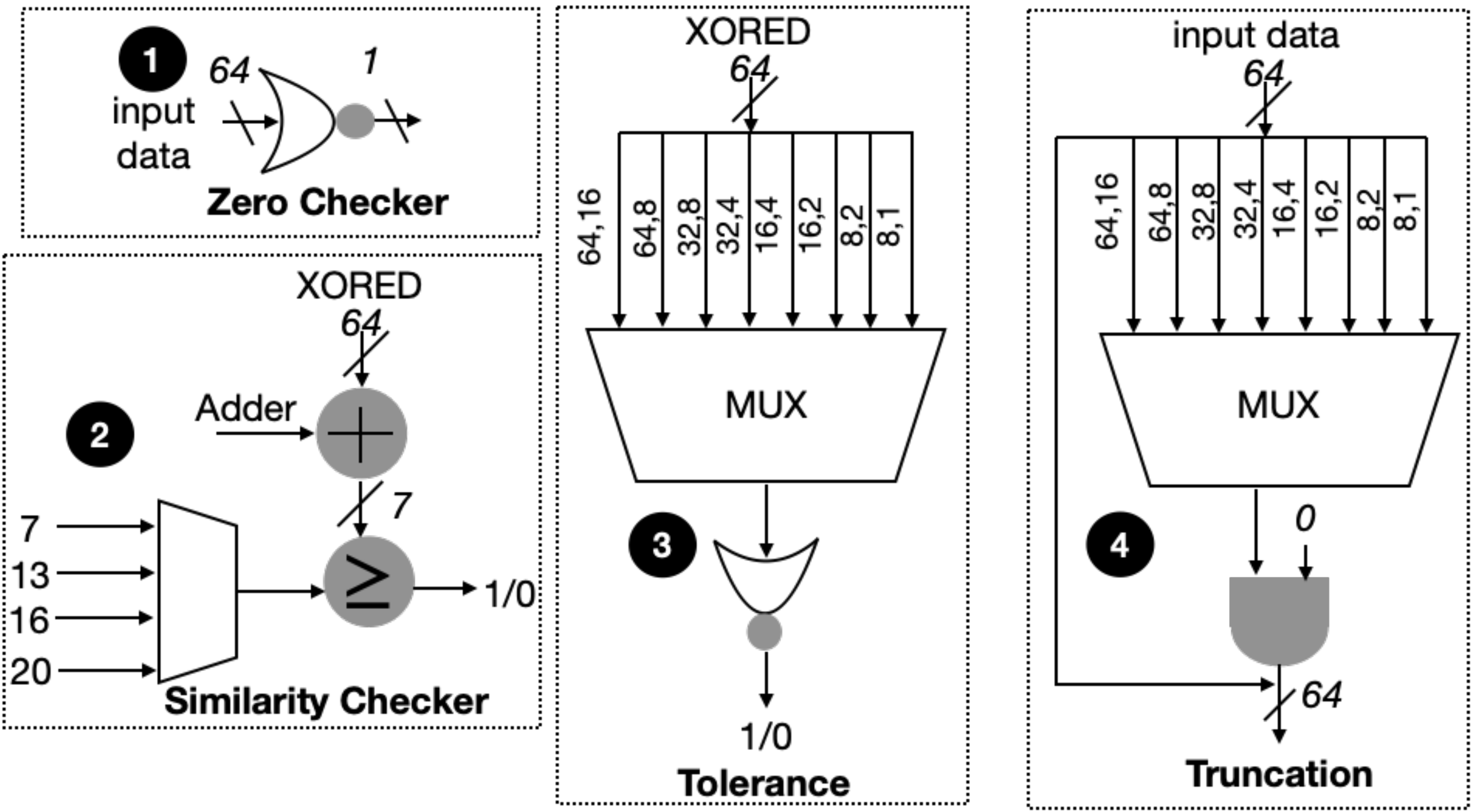}} &
\subfloat[]{\label{cdest}\includegraphics[height = 3.5cm, width =4cm]{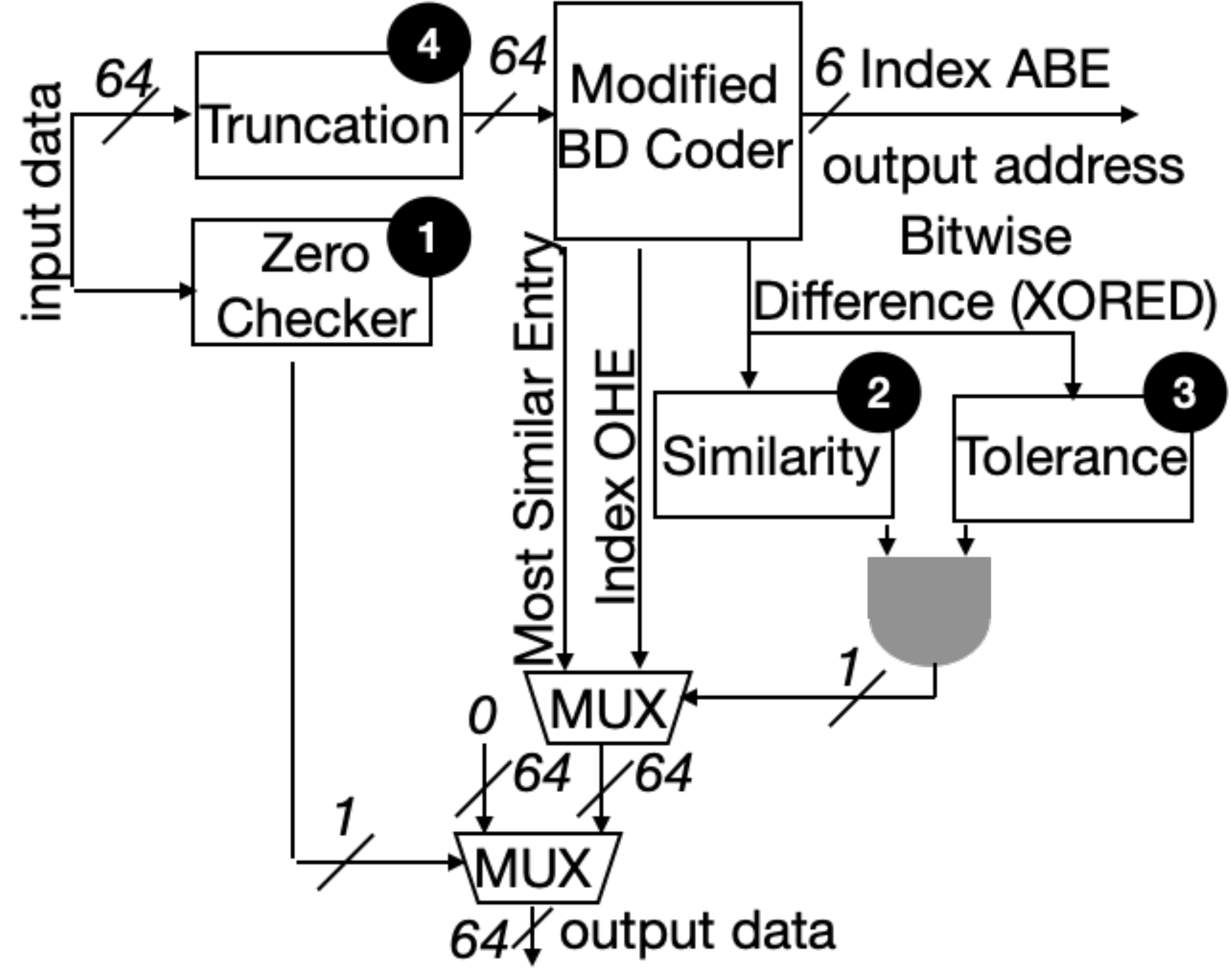}} &
\end{tabular}
\caption {\protect\subref{csub} Sub Modules for ZAC-DEST \protect\subref{cdest} ZAC-DEST Encoder Circuit}
\label{eval}
\end{figure*}
An additional row called the replica cell row is used in BD-Coder to count the number of 1's in the input data. The replica row will be used when the number of 1's is lesser than the MSE. We modify this row similarly as shown in Fig.~\ref{hamline}. The overall structure is shown in Fig.~\ref{cbdcoder} is called the Modified BD-Coder (MBDC).
\subsubsection*{\underline{Zero Checker}}
The zero checker circuit is used to detect, in advance, if the input data to be sent is all 0's. The zero checker gives an output 1, only when all the 64 input bits are 0's. 
This is achieved using the NOR gate as shown in Fig.~\ref{eval}~(\circled{1}). 
\subsubsection*{\underline{Similarity Checker}}
The similarity checker is shown in Fig.~\ref{eval}~(\circled{2}). It sums up the count of the number of dissimilar bits between input data and the most similar data. Depending upon the required similarity percentage, i.e. the percentage of bits which are supposed to be equal, 90\%, 80\%, 75\% and 70\% the numbers of dissimilar bits, irrespective of the bit positions, can be 7, 13, 16 and 20 respectively for 64-bit data. 
Hence, if an application requires a 90\% similarity, the sum of the bitwise difference should be less than 7 for ZAC-DEST and so on.  
\begin{figure}
\centering
\includegraphics[width=0.9\linewidth]{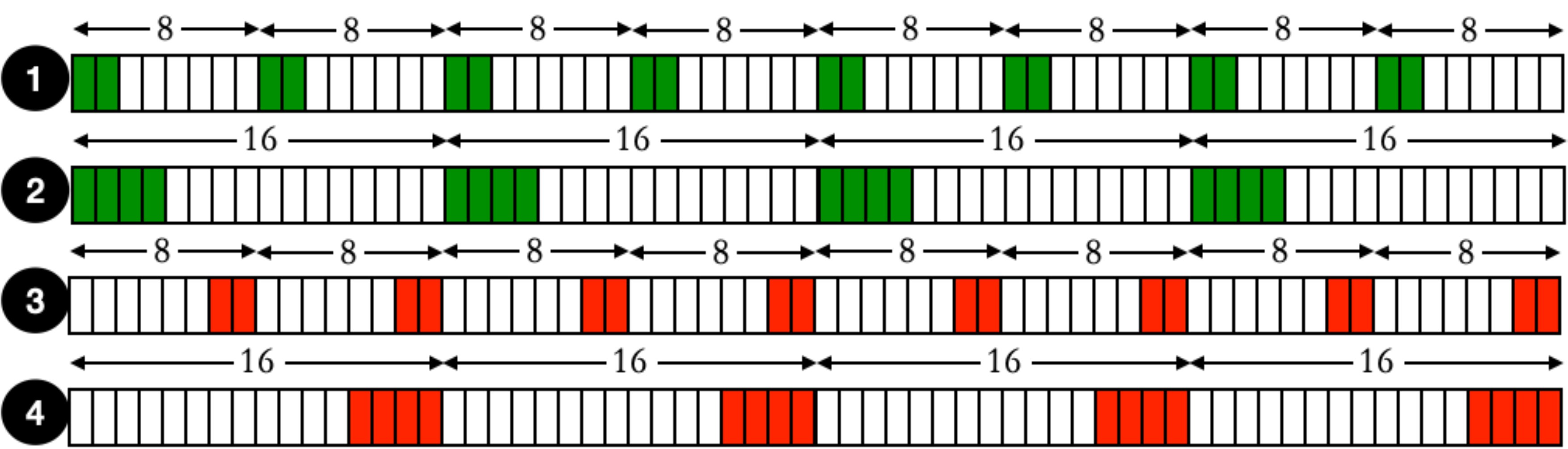}
\caption{Bits for Tolerance and Truncation 8, 2 and 16, 4. (Chunk Size = 8, 16, Bits within Chunk = 2, 4)}
\label{tt}
\end{figure}
\subsubsection*{\underline{Tolerance}}
The circuit design for introducing tolerance is shown in Fig.~\ref{eval}~(\circled{3}). At a time, we can transfer 64-bits of data. If we assume that this data contains eight chunks of 8-bit values, then tolerance will be applied to the MSB of each chunk. For a tolerance of 16, 2bit MSBs from each chunk cannot be approximated as shown in Fig.~\ref{tt}~(\circled{1}). Similarly if the values were of 16-bit each, i.e. there will be 4 chunks, then 4-bit MSBs of each of the chunks cannot be approximated as shown in Fig.~\ref{tt}~(\circled{2}).

Mostly the most significant bits (MSBs) are the ones that cannot tolerate approximation as described in Section~\ref{sec_evalall}. 
So, for 64-bit data, ZAC-DEST allows for the introduction of tolerance in 8 or 16 bit granularities. Depending upon the bit-width the tolerance bits can be distributed. The tolerance bits can be selected as per need using MUX as shown in fig.~\ref{csub}~(\circled{3}). For a bit-width of N, ZAC-DEST can have tolerance in first N/4 or N/8 MSBs, where N can have values of 8, 16, 32 and 64. A single mismatch (\emph{between data and the most similar entry}) in the tolerant bits will make the NOR gate output go low so that ZAC-DEST encoding is not applied and exact data is sent as shown in Fig.~(\ref{eval}).  
\subsubsection*{\underline{Truncation}}
The circuit design for introducing truncation is shown in Fig.~\ref{eval}~(\circled{4}). Similar to tolerance we allow support for a various bit-widths. 
ZAC-DEST allows a choice of N/4 and N/8 bit truncation for N equal to 8, 16, 32 and 64. The crucial difference is that truncation will make the bits to go to 0. For truncation of 16 and two different chunk sizes of 8 and 16, how the bits are approximated is shown in Fig.~\ref{tt}~(\circled{3}) and~(\circled{4}). 
\subsubsection*{\underline{Overall ZAC-DEST Encoding Scheme}}
The block diagram of ZAC-DEST encoder is shown in Fig.~\ref{cdest}. The input data to be sent over the channel is sent to zero checker. If the data is all 0's the zero checker output is 1 and all 0's are sent over the channel. At the receiver's end, all 0's are identified as such. The data is then forwarded to the MBDC to obtain MSE.
MBDC also provides the One Hot Encoded (OHE) address and the Address Binary Encoded (ABE) address of the most similar data. The most similar data is XORed with the original data to get the bitwise difference, which is then provided as input to the similarity and the tolerance blocks. The similarity block checks for the required similarity criteria and will output a 1 if the criteria is satisfied. The tolerance checker will output a 1 only if \emph{all} the bit positions selected for tolerance do not have a mismatch. If both similarity and tolerance criteria meet  
(ANDed output is 1), the One Hot Encoded Address is sent over the data lines (\emph{ZAC-DEST Output}), else the MSE is sent (\emph{BD-Coder Output}). Not if the original data has a lesser hamming weight that the MSE, the MBDC output the original data in place of MSE. A bit that informs the receiver whether the bits on the data lines represent the data or address. BD-Coder uses a single index line per chip to transfer the address. Since data table size is 64, a maximum of 6-bits are required to address the entire data table. The final output is sent after applying DBI.

\subsubsection*{\underline{MBDC Overheads}}
We have derived energy values of BD-Coder in 65nm to be 7~pJ from~\cite{seol2016energy}. The modification introduced in the data table is a single transistor that does not increase the energy significantly. We implemented the additional modules for ZAC-DEST in Verilog. We used 10,000 random inputs to generate the switching activity file  (SAIF) using Synopsys VCS tool. We used the SAIF file generated in Synopsys Design Compiler to obtain the power consumption of the hardware. The energy consumption overhead of the entire sub module is 9\% higher than that of the BD-Coder. The ZAC-DEST submodules, combined with BD-Coder consume 7.66~pJ per access. The latency for the data table in BD-Coder was 2.4~ns, while the entire ZAC-DEST sub module combined with BD-Coder has a latency of 3.4~ns. Even though the latency of MBDC increases as compared to BD-Coder, this is minimal as compared to the DRAM latency as also shown in~\cite{seol2016energy}. 
The area overheads of the submodules are 15\% higher as compared to the BD-Coder.
The receiver of both ZAC-DEST and BD-Coder is similar. Thus, the energy consumption, latency, and area of ZAC-DEST receiver is similar to that of BD-Coder's receiver. {These overheads are per DRAM chip, but overall overheads are still negligible as compared to DRAM as also shown in~\cite{seol2016energy}}. 

\section{Methodology}\label{sec:methodology}
\begin{table}[h]
\caption{Encoding Schemes Under Evaluation}
\centering
\begin{tabular}{ll}
\hline
OHE & One-Hot Encoding of ZAC-DEST\\
BDE\_ORG & Original Bitwise Difference Coder\\
BDE & Modified Bitwise Difference Coder\\
DBI & Dynamic Bus Inversion \\
ORG & Original Unencoded Data (Baseline) \\
\hline
\end{tabular}
\label{tab:encode}
\end{table}
ZAC-DEST improves channel-energy efficiency by transmitting approximate data in error resilient applications. Therefore we must choose a set of workloads that are amenable to approximation and have a quantifiable metric for measuring their output's quality. In this section, we describe the methodology used to evaluate the benefits of ZAC-DEST over existing models and the measure of quality used to understand the effect it has on the outputs.  
{Their analysis is done by first converting their inputs to hexadecimal traces. We then emulate the transfer of data over the DRAM channels using these traces and use them to simulate the models described in Table.~\ref{tab:encode}. For ZAC-DEST models that involve approximating data accesses, we use the simulated traces to reconstruct approximate inputs that are used to run the workloads. This way, we compare the results of the workloads with the original input set and the reconstructed ones to get a measure of quality.}
\subsection{Workloads}\label{sec:workloads}
\begin{figure}[h]
\centering
\includegraphics[width=\linewidth]{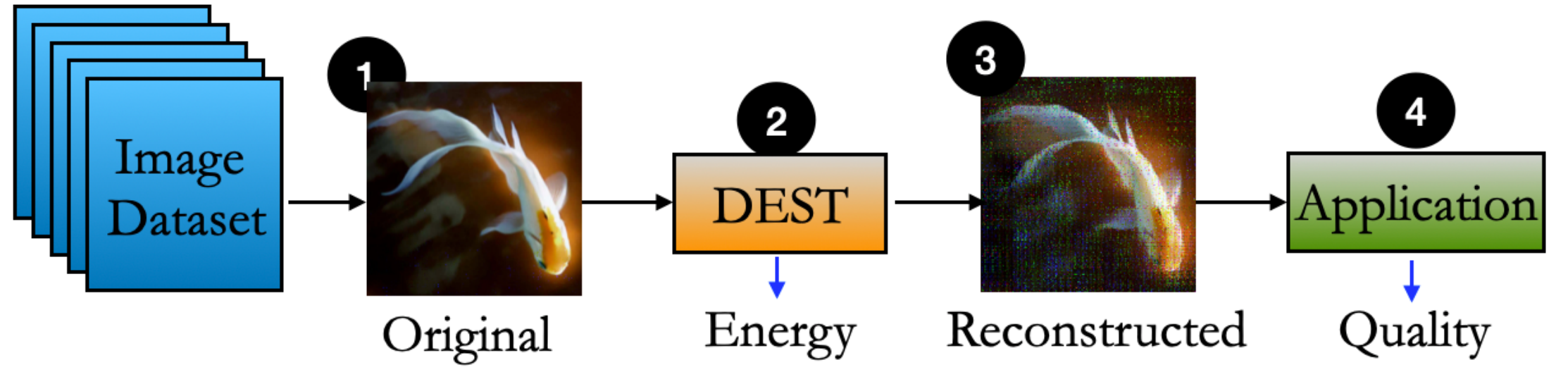}
\caption{Workflow of the methodology}
\label{fig:workflow}
\end{figure}

The workloads chosen for evaluation are machine learning applications that use images as inputs. To evaluate different models we Fig.~\ref{fig:workflow}~(\circled{1}) read the images and store their pixel values in a row-major format of 64 bytes chunks to simulate a cache line Fig.~\ref{fig:workflow}~(\circled{2}) apply ZAC-DEST and the other models on the resulting trace to simulate data transferred to the memory controller while calculating the amount of hamming and switching energy Fig.~\ref{fig:workflow}~(\circled{3}) reconstruct images using the data received by the memory controller Fig.~\ref{fig:workflow}~(\circled{4}) use the reconstructed images to run respective models and study the effect on quality. The workflow is summarized in Fig.~\ref{fig:workflow}.
Each workload has a different set of precision and accuracy metrics. Therefore we define \textit{quality} for each workload to understand the effect of ZAC-DEST on the output. \textit{Quality} is defined as the ratio of the output metric observed due to ZAC-DEST reconstructed images to that of the original images. As a result, a \textit{quality} of $1$ corresponds to the workload not expressing any degradation in its output and a \textit{quality} of 0.5 indicates the workload experiencing a 50\% degradation in the corresponding quality metric when compared to its non-approximated run. We now discuss each application in detail.

\subsubsection{ImageNet: CNNs from the ImageNet Challenge} 
Convolutional neural networks (CNNs) have been successfully applied in several image processing and computer vision tasks like image classification, object detection, etc~\cite{krizhevsky2014one, huang2017densely, szegedy2016rethinking, he2016deep, iandola2016squeezenet, simonyan2014very}. 
We use pre-trained pytorch~\cite{pytorch} models of 15 of these CNNs. These 15 CNNs were trained using the ImageNet 2012 classification dataset \cite{imagenet} which contains 1.28 million images in the \emph{training set.} We performed inferencing using 50K images in the \emph{validation set} of ImageNet dataset. The \emph{top-1} score matches the result with the highest probability against the target label. It is calculated as the number of times the top predicted label matches the target label, divided by the number of images evaluated.

\textbf{Quality Metric:} For these CNNs, the quality metric is a ratio of the \emph{top-1 score} for inferencing with ZAC-DEST reconstructed images and the original images.

\subsubsection{ResNet: Classification of the CIFAR dataset}
Previous works \cite{jacob2018quantization, szegedy2013intriguing} have shown that training ML models on approximate data are instrumental in alleviating drops in quality which accompany the use of approximate data. We demonstrate this by allowing ResNet-110 \cite{cifar}, a PyTorch model from the ImageNet challenge, to be trained on ZAC-DEST reconstructed train images before recording its accuracy while inferencing using ZAC-DEST reconstructed test images. We carry out these experiments on the CIFAR-100 dataset \cite{krizhevsky2009learning}. 

\textbf{Quality Metric:} It is a ratio of the \textit{top-1} score that we obtain from making predictions using reconstructed images (on the model that has been trained using reconstructed images) to that of the original data and model. 

\subsubsection{Quant: Color Quantization using K-Means}
Considering that both \textit{ResNet} and \textit{ImageNet} consist of Neural Networks, we chose \textit{Quant} as a workload for unsupervised tasks. This workload uses K-Means clustering to reduce the number of colours required to reproduce an image \cite{quant}. The algorithm reduces the large number of unique RGB values that are present in an image to a mere 64 with minimal degradation in image quality. This degradation is measured using the structural similarity (SSIM) \cite{wang2004image} metric that quantifies image quality degradation with respect to the reference image. We use the images from the KODAK image dataset \cite{kodak} and quantize the colour using \textit{Scikit-Learn's KMeans} algorithm in Python.

\textbf{Quality Metric:} It is a ratio of SSIM obtained using reconstructed images compared to the original images.

\subsubsection{Eigen: Using Eigen Vectors for face detection}
\textit{Eigen} is an unsupervised workload that uses Principal Component Analysis (PCA). PCA is a statistical procedure that uses transformations to convert a set of data into a set of uncorrelated variables. The task in this workload is to use PCA to decompose images present in the Yale Face Database \cite{yale} and then use these images for detecting faces.

\textbf{Quality Metric:} It is a ratio of the number of faces correctly detected using the reconstructed images when compared to using original images.
 
\subsubsection{SVM: FMNIST image classification using Python}
To compare the different encoding schemes on a sparse data, we choose an SVM model that learns the Fashion MNIST dataset \cite{xiao2017fashion,wenthundersvm18}. A Support-vector machine (SVM) is a machine learning model that uses a kernel to project data into higher dimensions following which it tries to learn a hyperplane that separates them distinctly. We choose FMNIST as it has a large number of sparse accesses, a behaviour that is exhibited by a number of contemporary workloads \cite{lee2018reducing}.

\textbf{Quality Metric:} It is the ratio of the number of articles of clothing correctly classified obtained using the reconstructed images when compared to using the original image set.

\section{Evaluation}~\label{sec_evalall}

In the following section, we discuss the setup used to analyze the effects of ZAC-DEST and understand how the different parameters that are used to control ZAC-DEST's approximations affect the energy savings and output quality.

\subsection{Setup}\label{sec:setup}

{We use C++ scripts to parse memory traces and simulate ZAC-DEST, DBI and BDE. These scripts are used for the dual purpose of simulating data transmission over the DRAM channel and it being received by the controller. Simulating data transmission is used to record the hamming and switching counts that are used for the energy calculations. Simulating data received by the memory controller, on the other hand, is used for evaluating the effect of ZAC-DEST on the output quality of the workloads.} Quality as defined in Section.~\ref{sec:methodology} refers to the ratio of \emph{top-1} precision for \textit{ImageNet} and \textit{ResNet}, SSIM values for \textit{Quant} and accuracy of workload task for \textit{Eigen} and \textit{SVM} obtained using ZAC-DEST reconstructed images when compared to using the original images.
The analysis for termination and switching energy is done as described in Section~\ref{sec:intro} and ~\ref{sec:background}. These values are calculated based on the data transmitted over the data lines and the index/other metadata passed over the control lines. While presenting the results, we discuss the termination energy, as in most cases both termination and switching follow similar trends.

We perform experiments for 8 chip DRAMs, with each chip having a data table size of 64. The choice of the data table size is made based on the discussions in~\cite{seol2016energy} where data table size up to 64 give a relatively large increase in energy benefits. 
\begin{figure}
\centering
\includegraphics[width=0.9\linewidth]{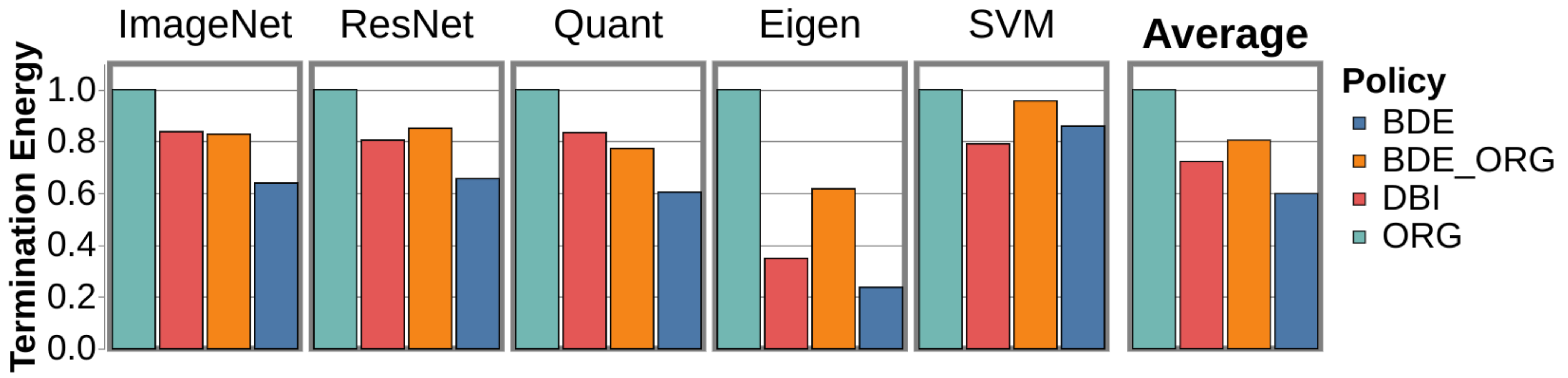}
\caption{Energy savings seen by all exact models}
\label{fig:allWorkloads_bde}
\end{figure}
\subsection{Comparing ORG, DBI, BDE\_ORG and BDE}
Fig.~\ref{fig:allWorkloads_bde} shows a comparison of the savings for all the exact models, i.e., non-approximate models, observed when compared to the original non-encoded scheme. We observe that when encoded with DBI, the number of 1’s being sent over the DRAM channel is reduced by 28\%, which leads to a corresponding decrease in termination energy when compared to original memory accesses. It is interesting to observe that data encoded using BDE\_ORG (proposed in~\cite{seol2016energy}) performs worse than DBI in this aspect leading to only a 20\% reduction while BDE with our proposed optimisations leads to a 41\% reduction. We hypothesize that this occurs due to the data tables not being updated regularly, thus leading to suboptimal encodings. Also, the overheads of transferring the address of the index adds up to the termination energy. Due to this, workloads like Eigen which use images that are relatively uniform suffer the most - observing only a 39\% reduction compared to 77\% reduction produced by our version of BDE that updates the data table at every access. Hence, for the remainder of this section, we compare the different modes of ZAC-DEST with respect to our modifed BDE, which acts as a stricter baseline.
\begin{figure*}
\centering
\includegraphics[width=\linewidth]{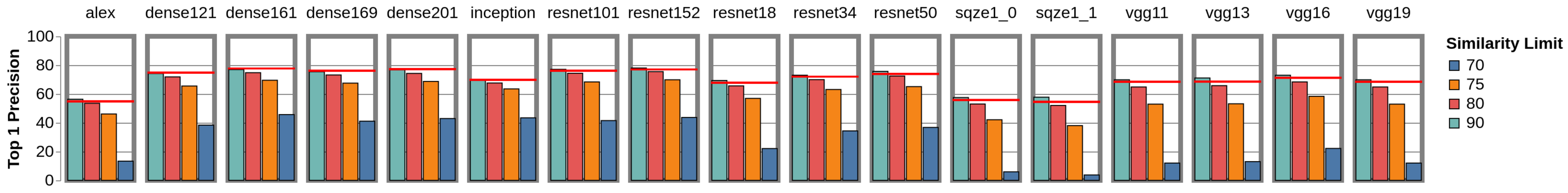}
\caption{Effect of Similarity Limit on \emph{top-1} precision for neural nets in the ImageNet Challenge. The red line denotes the original accuracy}
\label{fig:nn_quality_sim}
\end{figure*}
\begin{figure}
\begin{center}
\subfloat[Limit = 90]{\includegraphics[width=0.25\linewidth]{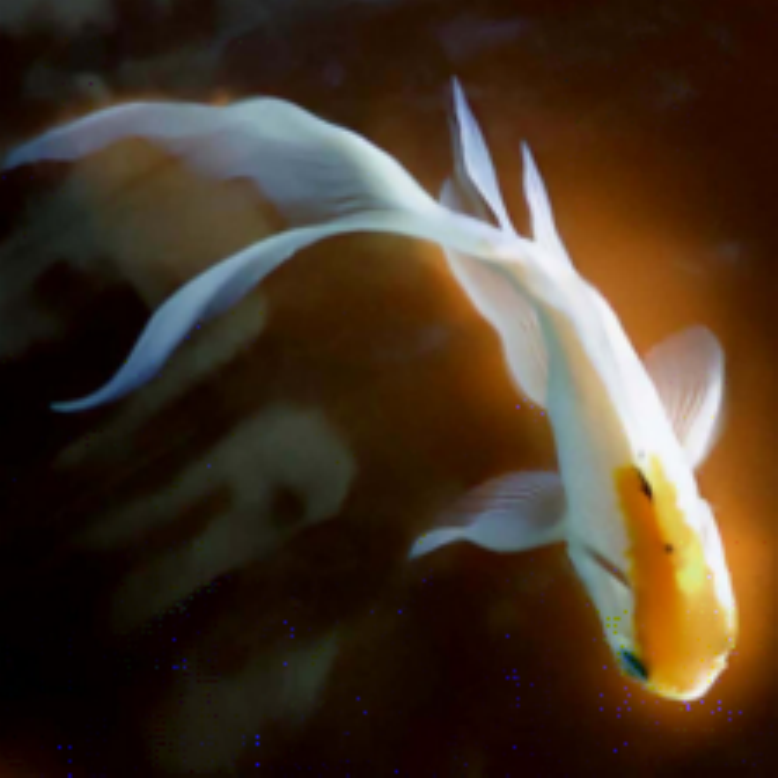}}\
\subfloat[Limit = 80]{\includegraphics[width=0.25\linewidth]{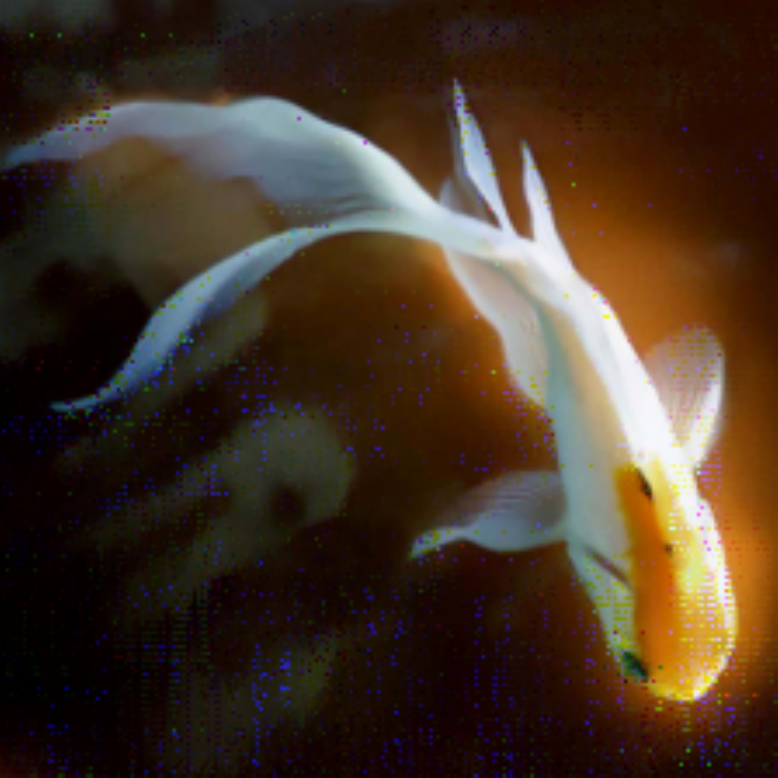}} \
\subfloat[Limit = 70]{\includegraphics[width=0.25\linewidth]{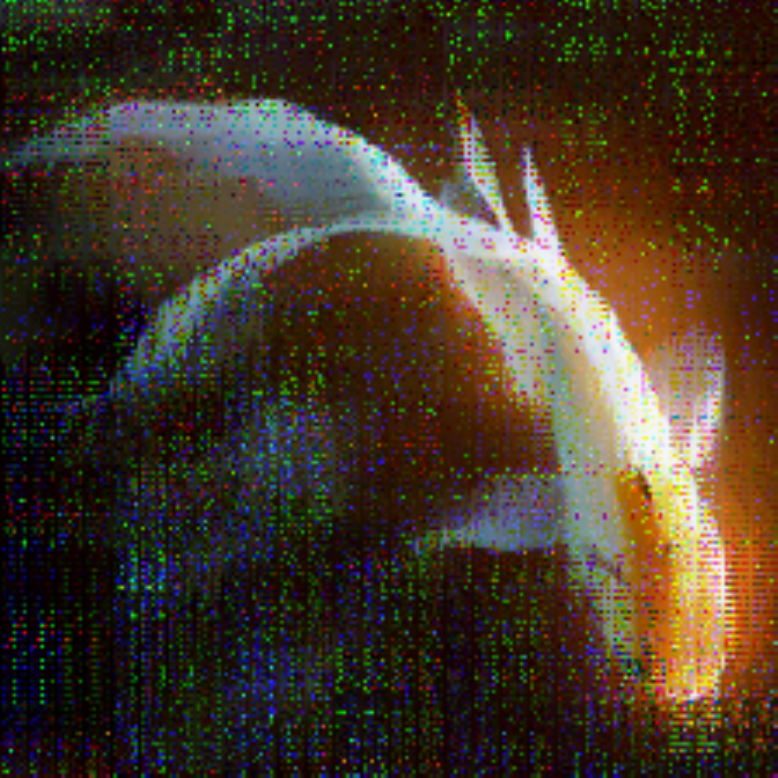}}\
\end{center}
\caption{Reconstructed images for different Similarity Limits}
\label{fig:sim}
\end{figure}
\subsection{Effect of Similarity Limit}\label{sec:simil}
The \textit{Similarity Limit} is a parameter that controls the amount of approximation being done to the workload. A \textit{Similarity Limit} of 90 denotes a ZAC-DEST implementation where data accesses at least 90\% of bits similar to the most similar entry would be approximated. We choose 90\%, 80\%, 75\% and
70\% (these correspond to a max of 7, 13, 16 and 20 bits being approximated) as similarity limits for analysis as they provide a varied view of the benefits that the approximation can yield. Allowing for
more bits to be approximated (a similarity limit of
$<$ 60\%) would lead to incorrect results while
high thresholds (a limit of $>$90 \%) would not result in
any significant improvement in energy savings. For these experiments, both \textit{Truncation} and \textit{Tolerance} are kept as $0$.
 Fig.~\ref{fig:nn_quality_sim} shows the behaviour of the CNNs from the ImageNet Challenge. There is a decline in \emph{top-1} precision as we decrease the \textit{Limit} due to loss in image quality. It is interesting to observe that the loss in accuracy in decreasing the \textit{Limit} from 75 to 70 is much more significant than the other transitions, namely from 90 to 80 and 80 to 75. Fig.~\ref{fig:sim} shows the degradation of the reconstructed image caused due to the decrease in \textit{Similarity Limit}.
\begin{figure}
\centering
\includegraphics[width=\linewidth]{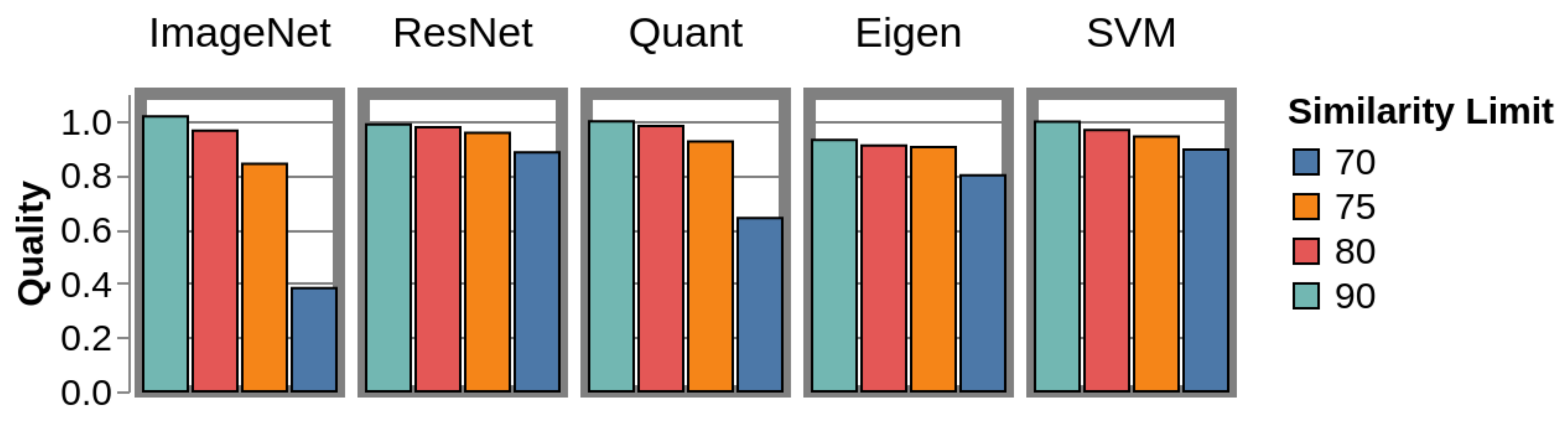}
\caption{Effect of Similarity Limit on output quality for all workloads.}
\label{fig:allWorkloads_quality_sim}
\end{figure}
\begin{figure}
\centering
\includegraphics[width=\linewidth]{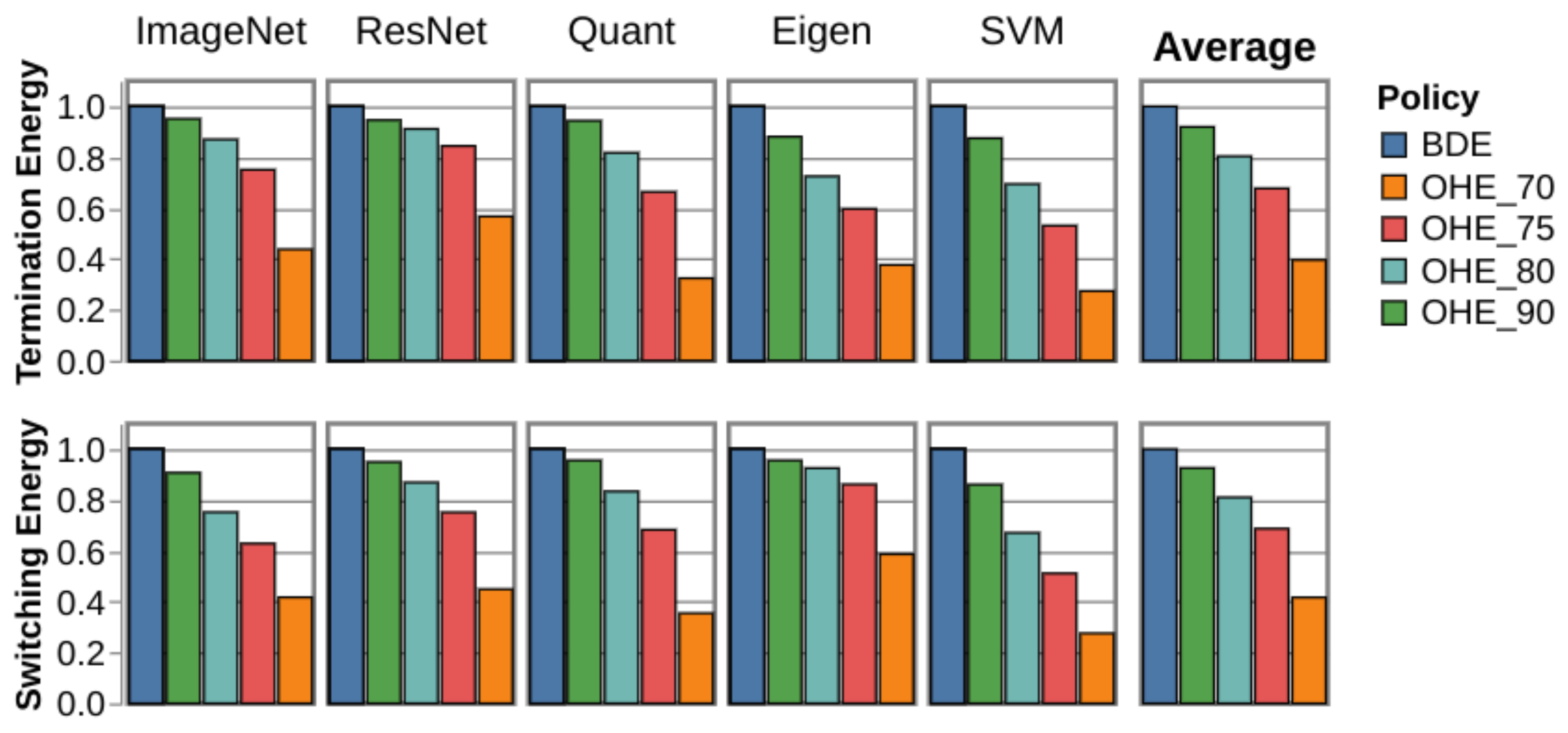}
\caption{Energy savings observed by all models with ZAC-DEST while varying its similarity limit.}
\label{fig:allWorkloads_sim}
\end{figure}
When we compare quality metrics across workloads in Fig.~\ref{fig:allWorkloads_quality_sim}, we observe a similar trend of decreasing qualities with a decrease in \textit{Similarity Limit}. While in the case of the Eigen, ResNet and SVM, it is gradual, ImageNet and Quant observe a sharper decline as the Limit decreases. It is important to note that for a \textit{Similarity Limit} of 90 most of the workloads have a quality comparable to or more than 1 (where a quality of 1 means that there is no reduction in accuracy). Fig.~\ref{fig:allWorkloads_sim}, shows the effect of \textit{Similarity Limit} on termination and switching energy for all the workloads. We observe that for a similarity limit of 90, as compared to BDE, ZAC-DEST reduces the termination and switching energies by 8\% and 7\% respectively. Decreasing the similarity limit (allowing more bits to be approximated) drastically reduces energy consumption. Comparing the energy consumption for \textit{Similarity Limits} 90~/~80~/~75~/~70, we observe a reduction of 8\% / 20\% / 32\% / 60\% in termination energy compared to BDE, with a similar trend for switching energy. These are especially promising results as for \textit{Similarity Limit} of 80 and 75 we see a reduction of 20\% / 32\% in the energy consumption when compared to BDE with qualities of 0.96/0.8.
\subsection{Effect of Truncation and Tolerance}
\begin{figure*}
\centering
\includegraphics[width=0.95\linewidth]{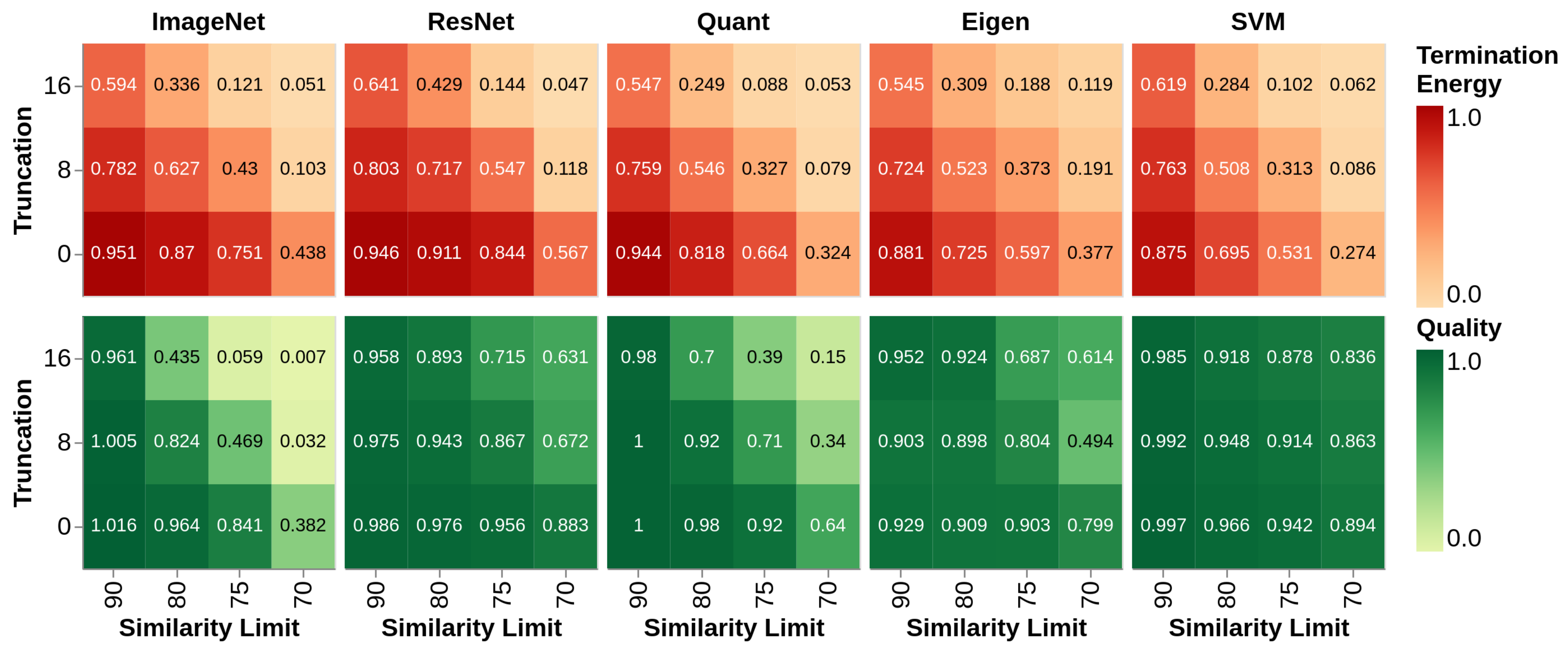}
\caption{Effect of Truncation and Similarity Limit on Termination Energy and Quality (Switching Energy follows similar trends). Each number in the box is the value of the metric.}
\label{fig:all_trunc_heat}
\end{figure*}

Fig.~\ref{fig:all_trunc_heat}, shows the effect of \textit{Truncation} and \textit{Similarity Limit} on the energy and quality of workloads. We observe that increasing \textit{ Truncation} results in a decrease in energy at cost of quality. This is caused due to the increase in the number of bits being masked to zero caused by increasing \textit{ Truncation}. For a \textit{Limit} of 80, increasing \textit{Truncation} from 0 to 16 causes the savings of both termination and switching energy to increase from 20\% to 68\% as compared to BDE. 
But at the same time, we observe the quality to drop from 0.96 to 0.77. It is interesting to observe that the effect of Truncation becomes more prominent on lower Similarity Limits, with a drop in quality from 0.72 to 0.44  for a \textit{Limit}~of~70. 
\begin{figure*}
\centering
\includegraphics[width=\linewidth]{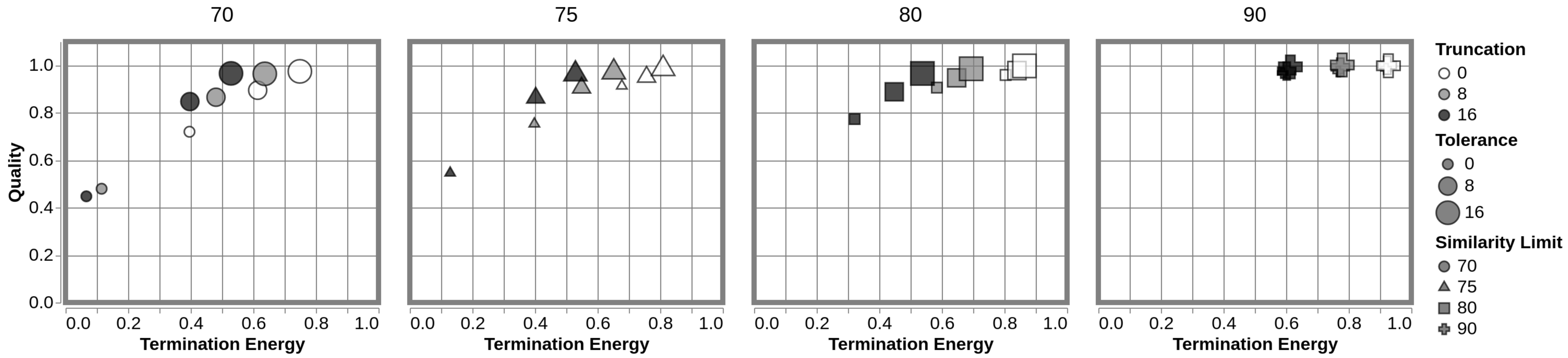}
\caption{Effect of ZAC-DEST on Quality and Energy as an average over all workloads. Darker points correspond to higher Truncation, larger points correspond to larger tolerance, and more number of sides correspond to larger similarity limits.}
\label{fig:all_scatter}
\end{figure*}
Fig.~\ref{fig:all_scatter}, shows the effects of different parameters on workloads. Each data point is differentiated based on color, size and shape which correspond to \textit{Truncation}, \textit{Tolerance} and \textit{Similarity Limit}, respectively. This plot helps visualize the combined effects that different parameters have on energy savings and quality degradation. Ideally, we would select parameters to minimize the energy consumption without compromising on quality, selecting design points on the \textit{top-left} of the chart. We observe that decreasing \textit{Limit} and increasing \textit{ Truncation} results in energy savings at the cost of quality, pushing design points to the \textit{lower left}. We use \textit{ Tolerance} to balance the effect of those parameters. Increasing it (represented by increasing the size of the point) restricts the number of times ZAC-DEST can be true, thus resulting in lower energy savings but better quality (pushes the design points to the \textit{top right}). Just as in the case of \textit{Truncation}, \textit{Tolerance} does not affect the quality and energy savings by a large amount at higher \textit{Similarity Limits} (where the design points of different sizes and colours are closer to each other), but increases as we lower the \textit{Limit}. 

\subsection{Using Reconstructed Images for Training}\label{sec:train}

\begin{figure}
\centering
\includegraphics[width=\linewidth]{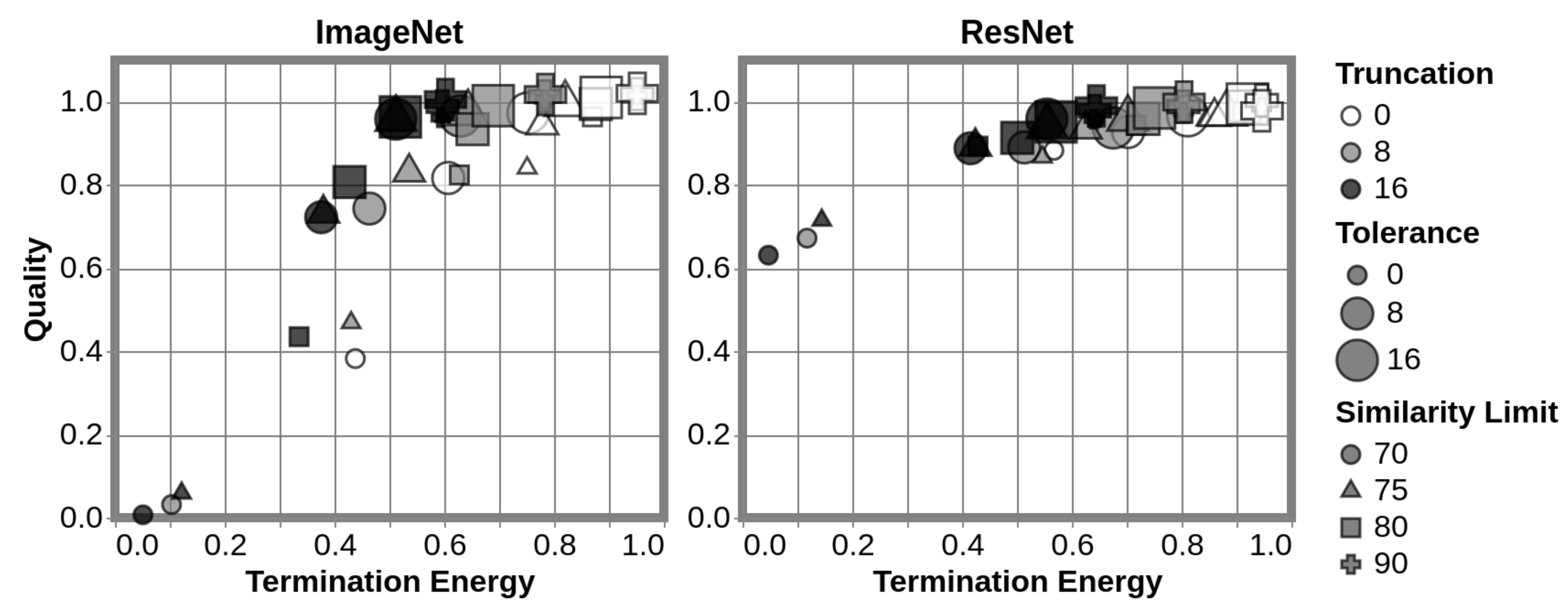}
\caption{Effect of ZAC-DEST on the ImageNet and ResNet}
\label{fig:resimg_scatter}
\end{figure}

\begin{figure}
\centering
\includegraphics[width=\linewidth]{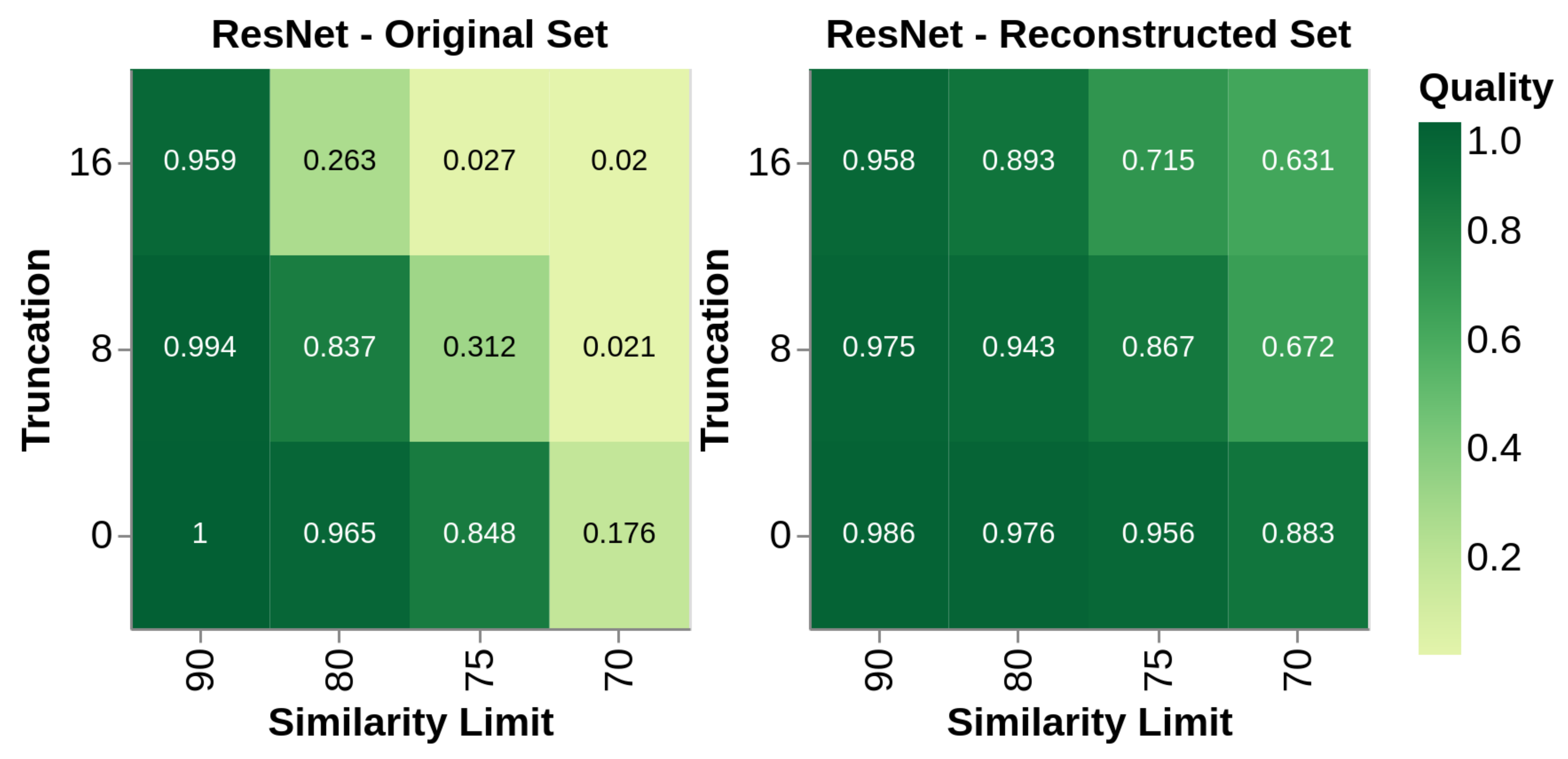}
\caption{Comparing ResNet-110 for different training sets}
\label{fig:cifar}
\end{figure}

The workload \textit{ResNet} is used to demonstrate that training models on images reconstructed using ZAC-DEST, i.e., on approximate images, would alleviate some of the quality degradations. We observe this behaviour when we compare \textit{ImageNet} and \textit{ResNet} in Fig.~\ref{fig:resimg_scatter}. We compare two different models of \textit{ResNet} - one that has been trained using the reconstructed images while the other has been trained using the original dataset. Fig.~\ref{fig:cifar} compares the quality of the two models based on the effects of \textit{Similarity Limit} and \textit{Truncation}. We observe that the drop in quality is smaller in the case of \textit{ResNet} trained on approximation images as compared to the model that isn't. {This motivates training models with the ZAC-DEST reconstructed data when feasible to  improve the accuracy of the application. In some configurations, we observe an improvement of up to 9$\times$ in output quality. Hence depending on the application, in case where higher accuracy is needed ZAC-DEST can be used both while training and inference. }
\subsection{Effect of ZAC-DEST on Output Quality}

Fig.~\ref{fig:all_trunc_heat} provides an insight into how amenable each workload is towards approximation. We observe that \textit{ResNet} and \textit{SVM} are more tolerant to higher levels of approximation compared to \textit{ImageNet} and \textit{Quant} despite observing similar benefits in energy consumption. Fig.~\ref{fig:resimg_scatter} shows this analysis for \textit{ImageNet} and \textit{ResNet} as representatives of the different behaviours. Here, \textit{ImageNet} dips sharply at higher approximation configurations while \textit{ResNet} manages to remain stable, i.e., it does not experience as large a drop in quality. This behaviour is directly related to the nature of each workload. For \textit{Quant} large variations in the image can cause the K-Means algorithm to quantize colours in a poor manner, leading to lower values of SSIM. \textit{SVM}, on the other hand, being a generally robust model classifying a relatively simple data set is amenable to approximations.

\begin{figure}
\centering
\includegraphics[width=\linewidth]{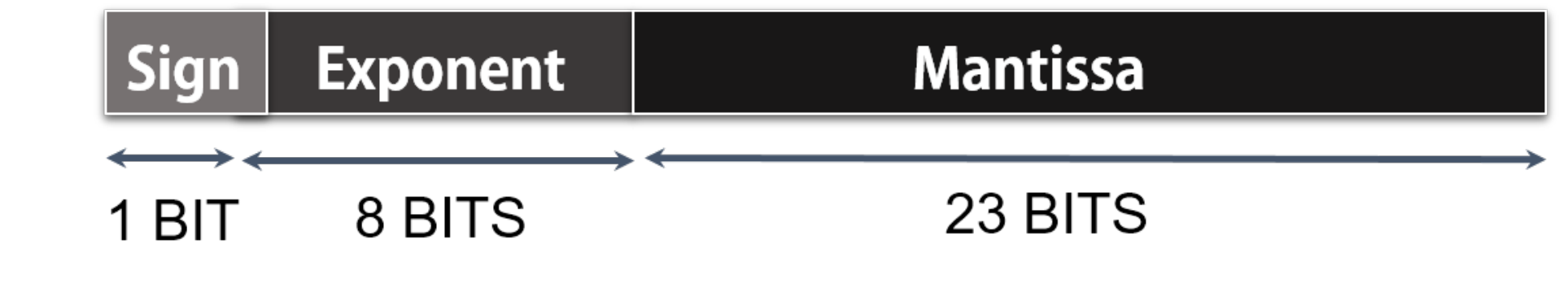}
\caption{32-bit Floating Point representation in IEEE 754}
\label{fig:ieee}
\end{figure}

\begin{figure}
\centering
\includegraphics[width=\linewidth]{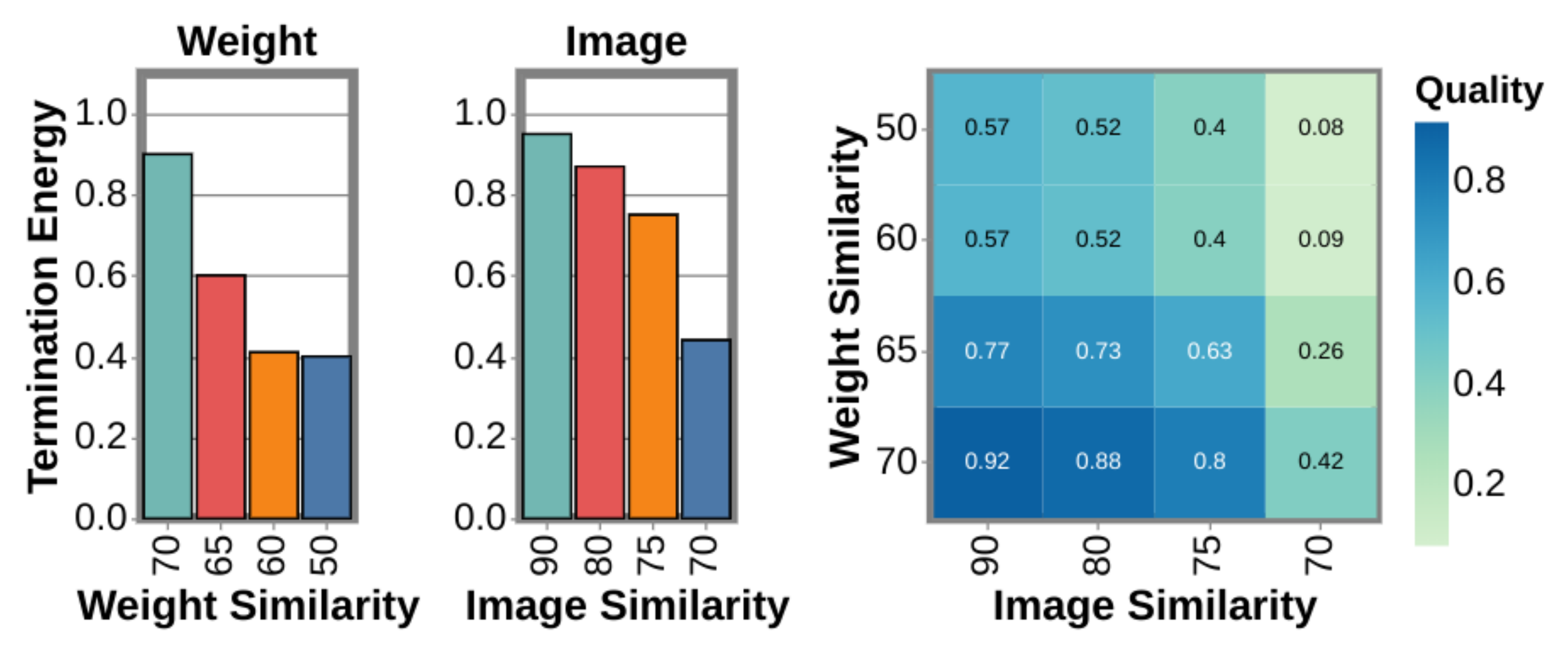}
\caption{Comparing InceptionNet for both weight and image approximation}
\label{fig:inception}
\end{figure}

\begin{figure}
\centering
\includegraphics[width=\linewidth]{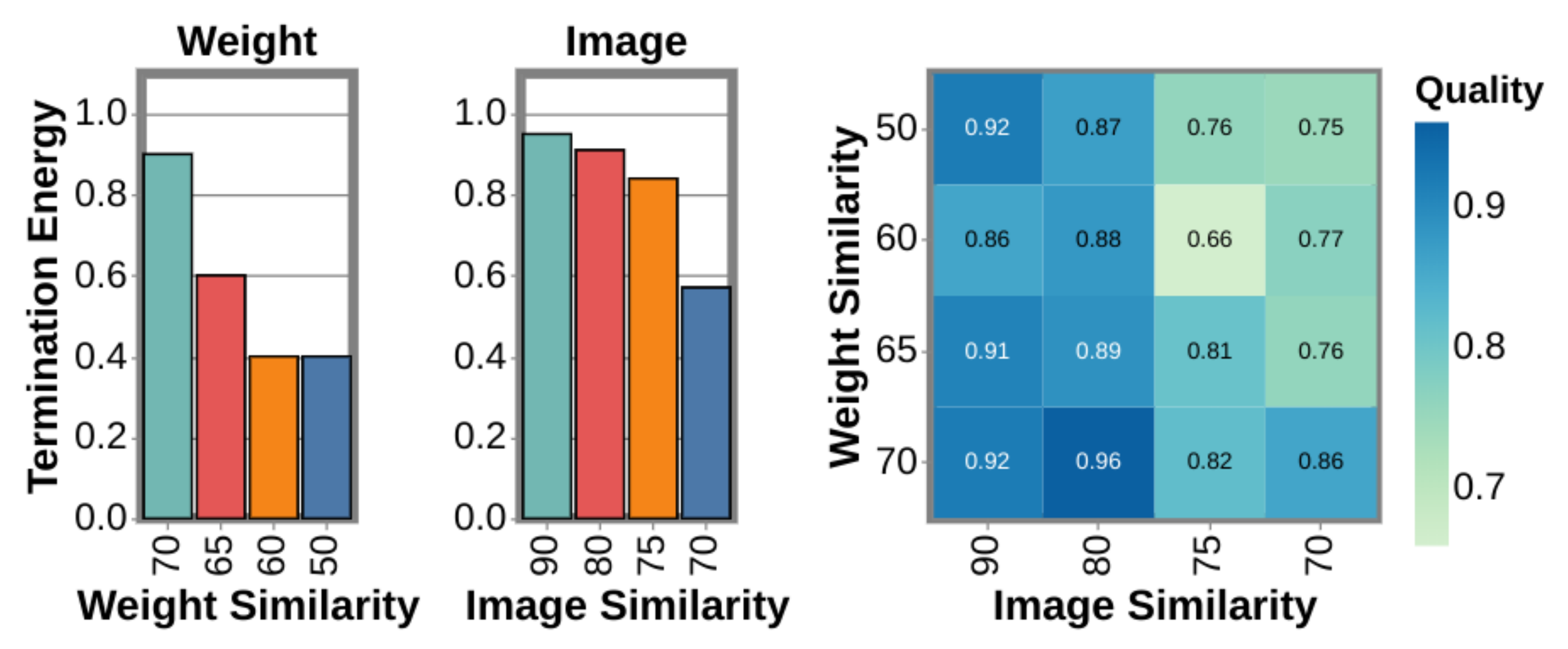}
\caption{Comparing ResNet-110 for both weight and image approximation with training}
\label{fig:resnet}
\end{figure}
\subsection{Effect of ZAC-DEST on Both Weights and Images}
    We now study the effect of applying ZAC-DEST on both the weights and the images to study the impact on energy and quality. For approximating weights we follow a similar strategy as approximating images. The weights are represented using the IEEE 754 format as shown in Fig.~\ref{fig:ieee}. It is important to note that for weights it is imperative that we do not approximate the exponent and sign bits as it introduces large errors into the calculations. We evaluated and observed that approximating even the last bit of exponent leads to 60\% deterioration in output quality. Thus, based on structure of the traces and the DRAM data layout (detailed in Fig~\ref{drammain}) we set the tolerance sign and exponents bits are not approximated.
Fig.~\ref{fig:inception}, shows us the effect of ZAC-DEST on termination energy and quality when both the images and weights for the model “InceptionNet” from the ImageNet workloads for varying \emph{Similarity Limits}. For \emph{Similarity Limits} 70 / 65 / 60 / 50, we observe a reduction of 10\% / 40\% / 59\% / 60\% in termination energy (due to weights) compared to BDE. We see that for such savings in energy the quality reduces from 0.92 to 0.57 (for a fixed image \emph{Similarity Limit} of 90\%). Fig.~\ref{fig:resnet}, shows us that the effect of ZAC-DEST on ResNet-110 when we approximate both weights and images during both training and testing. Similar to what was discussed in Section~\ref{sec:train}, we see that training with ZAC-DEST improves the output quality. Such comparisons would be useful for determining the correct modes to be used for different models to obtain the desired output quality. Based on the whether weight or image transfer dominates depending upon the hardware configurations and the application, for acceptable quality drops one among the variety of configuration can be chosen.
\subsection{Instances of Encoding During Memory Transfers}
\begin{figure}
\begin{center}
\subfloat[]{\includegraphics[width=0.7\linewidth]{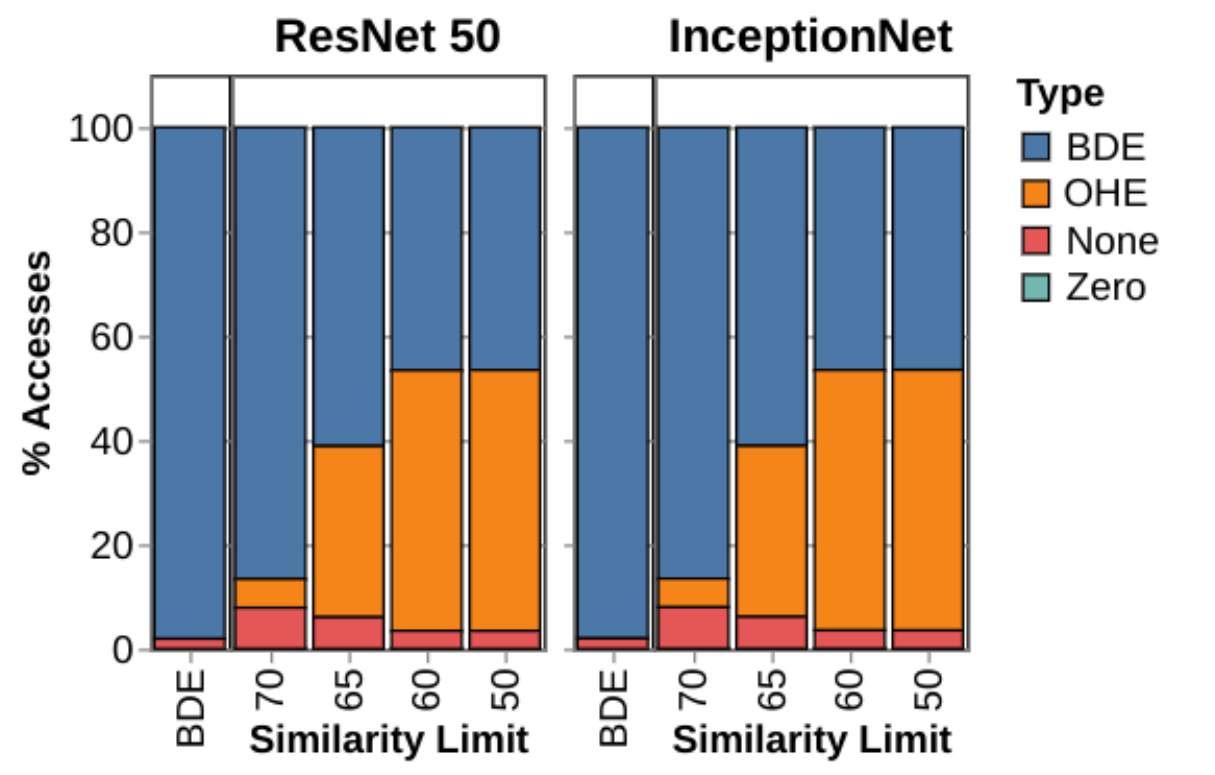}} \
\subfloat[]{\includegraphics[width=0.9\linewidth]{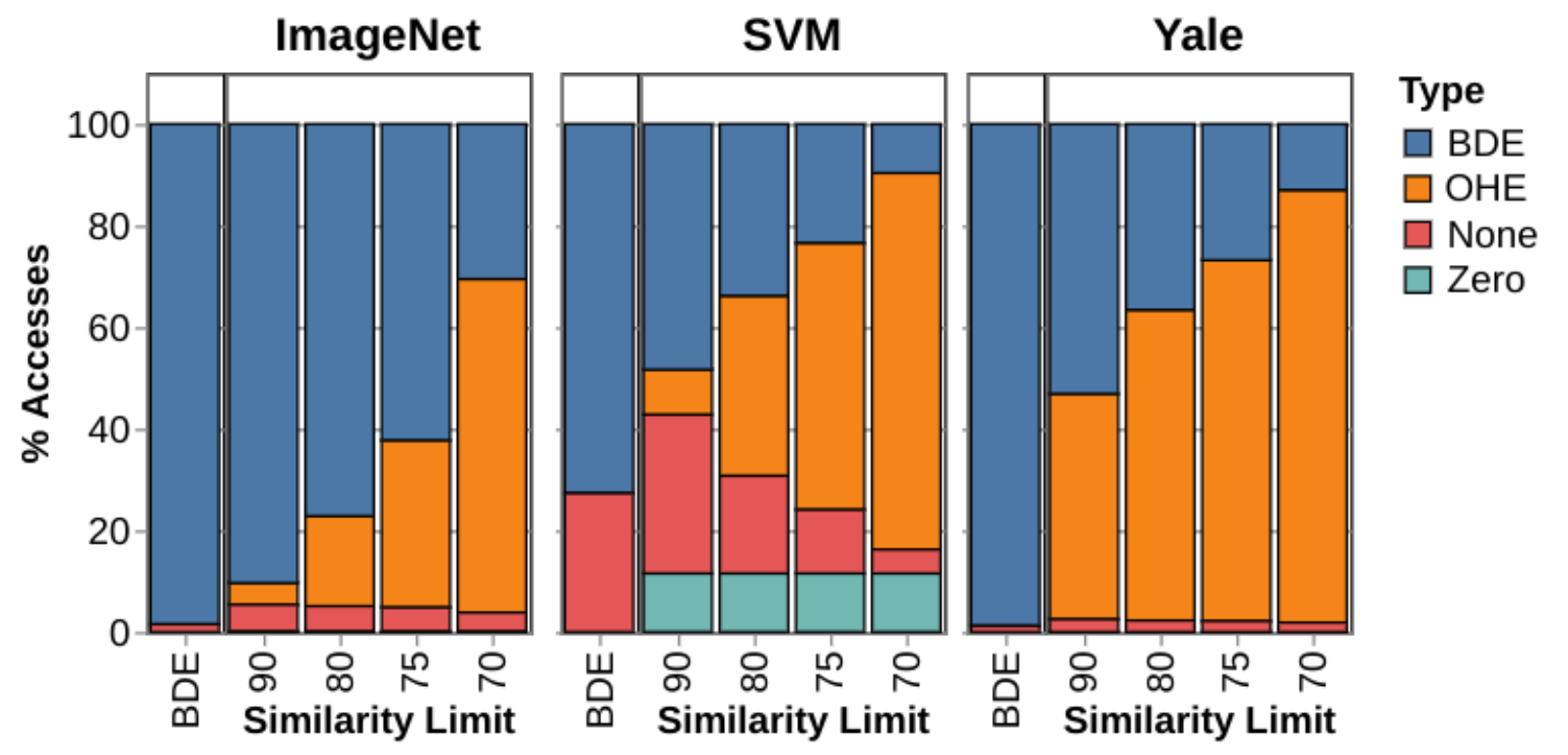}}\
\end{center}
\caption{Frequency with which data is encoded using ZAC-DEST and BDE during a) Weight Transfers b) Image Transfers for varying similarity limits}
\label{fig:inst_all}
\end{figure}
In Fig.~\ref{fig:inst_all} we visualize the frequency with which the data is encoded for a particular encoding scheme for both weights and images. We compare these values when we use BDE and ZAC-DEST (which is built over an optimized version of BDE). As compared to the BDE proposed in~\cite{seol2016energy}, we have added two modifications. i) We handle 0's separately, and ii) We have a stricter condition for BDE as we sum the hamming weight of both the data and index values to evaluate the BDE condition. In~\cite{seol2016energy}, only hamming weight of data is considered and not the index values. In both cases, we see that a majority of the accesses are encoded using either of the schemes, with only an average of 6.5\% and 6.6\% of the accesses not being encoded using ZAC-DEST and BDE respectively. This result demonstrates the high similarity between the transferred data and also speaks to the fact that to improve over BDE, whose coverage is already very high, it is imperative to implement schemes that have a better encoding mechanisms.

\section{Related Work}\label{sec:related}
In this section, we briefly present work in the area of data encoding/compression and approximation.

\subsubsection*{\underline{Data Encoding}}
ZAC-DEST is an approximate encoding model that develops on the state of the art data encoding schemes BD-Coder \cite{seol2016energy} and DBI \cite{stan1995bus} to give higher energy benefits. Various other works in the past have focused on energy reduction using data encoding in DRAM channels. Yan et al.~\cite{yang2004frequent} proposed an encoding scheme which exploits temporal locality of data words. It uses one-hot encoding to send frequently occurring values. Suresh et al.~\cite{suresh2005valve} proposed VALVE, a variable-length bit pattern for encoding and decoding. it matched partial data and sent either one-hot code or two-hot code masks for that partial data word while the rest was sent unencoded. Lee et al. proposed SILENT ~\cite{lee2004silent}, a data encoding scheme which focused on reducing the switching energy by reducing the hamming weight of the data words by exploiting similarity between current and the previously accessed data word. Lee et al.~\cite{lee2018reducing} proposed a data encoding scheme for GPUs, which reduces the number of 1's in the data sent over the channel. It took special care of zero data by encoding it with a constant with reduced hamming weight. While the technique works well for GPU applications it has been shown to perform poorly for CPU applications. 

Stanley-Marbell et al. \cite{vdms} propose a value-deviation-bounded serial (VDBS) approximate encoding scheme that significantly reduces the switching observed for data. Pekhimenko et al. \cite{toggel} propose Toggle-Aware Compression schemes that reduce switching count impact of the data compression algorithms. Both schemes can be used to assist in alleviating the increase in switching counts caused due to BDE in certain workloads (as seen in Fig.~\ref{fig:allWorkloads_bde}). 

\subsubsection*{\underline{Approximation in Hardware}}
Various works have focused on the introduction of approximation to DRAMs, caches and processors~\cite{koppula2019eden}. Sampson et al. \cite{sampson2014approximate,sampson2015accept,sampson2011enerj} have proposed frameworks for annotating and identifying regions in the program that are amenable to approximation and hardware mechanisms for memories that result in energy savings at the cost of output quality. Liu et al. \cite{liu2011flikker} use application-level input to effectively reduce the refresh rate of DRAMs, which may result in data corruption.  
Miguel et al. \cite{lva} proposed Load Value Approximation (LVA) and Thwaites et al. \cite{roll} proposed rollback-free value prediction, techniques that approximately predict the data to be accessed during a load. As such behaviour results in increased number of predictions being made and reduces the number of times the memory is accessed. These works focus on introducing approximation in a method that is different from ZAC-DEST, which makes it entirely possible to stack them with ZAC-DEST to leverage more benefits.

Miguel et al. \cite{dop} propose Doppelganger, an approximate cache mechanism that associates multiple similar entries together to reduce the amount of data stored. Boyapati et al. \cite{boyapati2017approx} propose APPROX-NoC, a mechanism for network-on-chip (NoC) devices to eliminate the transmission of similar cache blocks by encoding them to similar data patterns. Both these works can function in synergy with ZAC-DEST.

\section{Conclusion}\label{sec:conclusion}
In this paper we propose DEST, an approximate data encoding scheme to reduce DRAM channel energy consumption for error resilient applications. DEST works by exploiting data similarity and the error resilience of applications leading to reduction in hamming weight (number of 1's in data word). DEST builds up on top of existing data encoding schemes namely BD-Coder and DBI. We applied DEST on five different set of machine learning applications and observed a reduction of 40\% and 37\% in termination and switching energy respectively as compared to the state of the art data encoding technique with an average output quality loss of 10\%. DEST, if applied on both training and testing can significantly outperform designs that apply DEST only during testing, but are trained on non-DEST encoded data.

\section{Acknowledgements}
This work is supported through grants received from Intel, SMDP-C2SD and YFRF/PhD fellowship
under Visvesvaraya PhD scheme from the Ministry of Electronics and IT, and through SERB grants
CRG/2018/005013 and MTR/2019/001605. This work is partially supported through Ashoka University startup and Huawei Technologies India grants to Manu Awasthi. This work is also supported through grant received from SRC Grant  2980.001 and 2020-IR-3005.

\bibliographystyle{ieeetr}

\begin{IEEEbiography}[{\includegraphics[width=1.1in,height=1.25in]{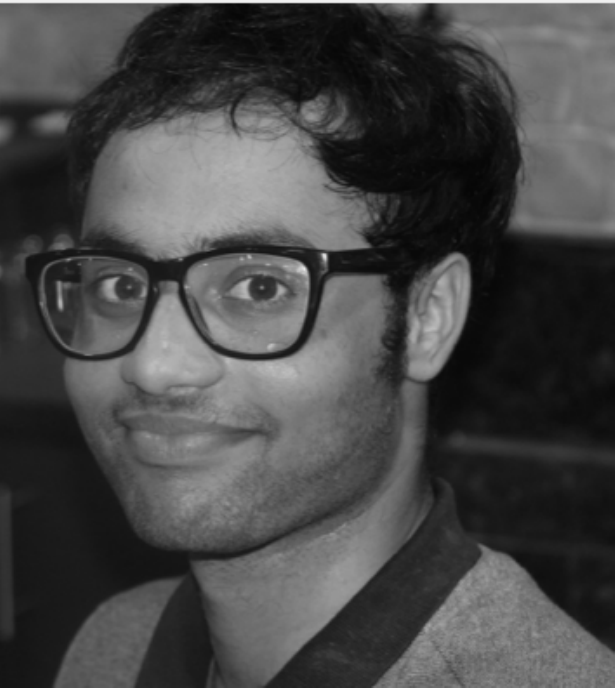}}]{Chandan Kumar Jha}
recieved his B.Tech
Degree from National Institute of Technology Meghalaya, Shillong, India in 2015. He is currently a PhD student in Electrical Engineering at Indian Institute of Technology Gandhinagar, India. His research interest include approximate circuits, approximate architectures and energy efficient systems design. He was the recipient of Merit Scholarship during his B.Tech. He was the recipient of Visvesvaraya PhD Fellowship from 2015 to 2019. He is currently an Intel PhD fellow from 2019 onwards. 
\end{IEEEbiography}

\begin{IEEEbiography}[{\includegraphics[width=1.1in,height=1.25in]{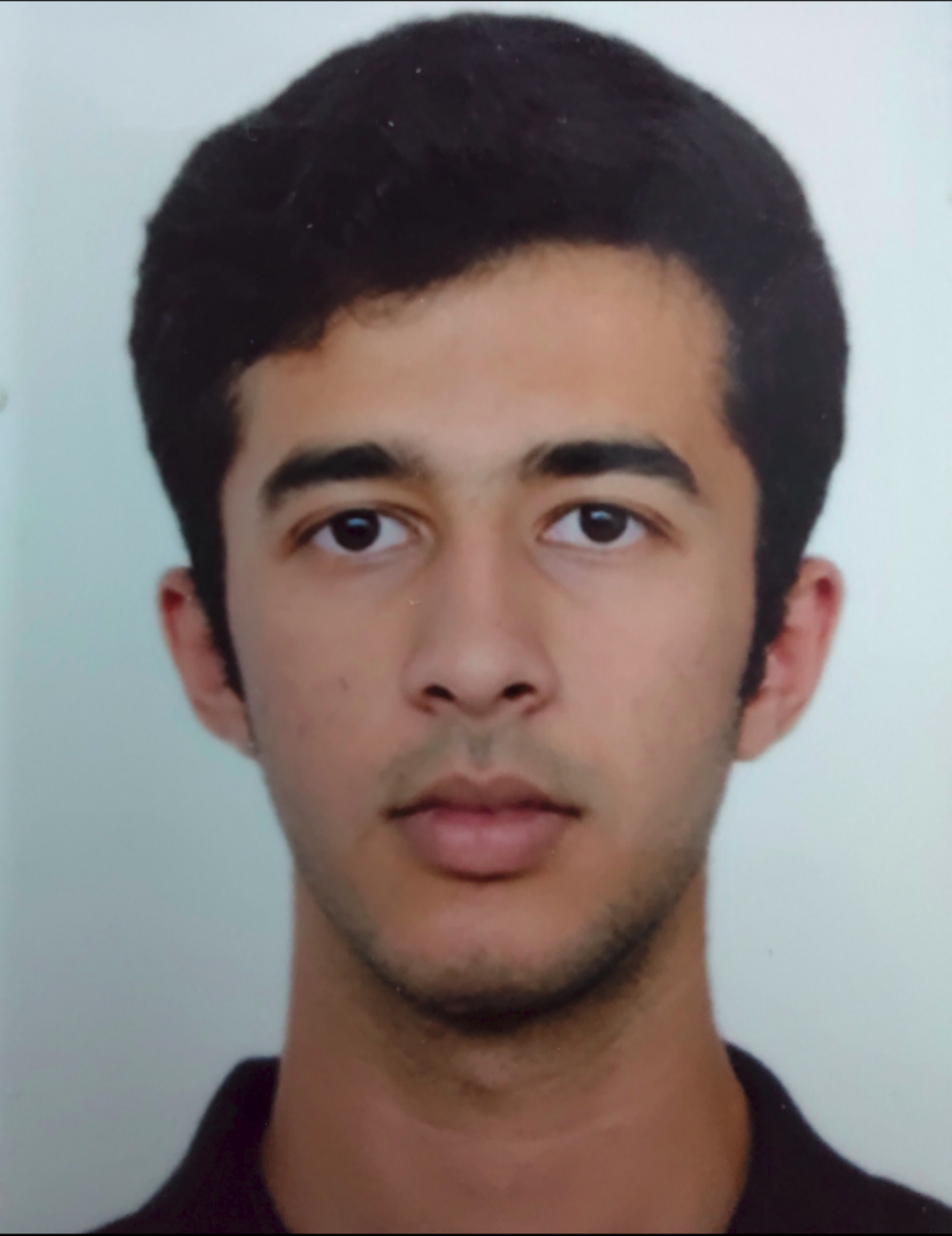}}]
{Shreyas Singh} received his B.Tech. Degree in Computer Science and Engineering from the Indian Institute of Technology, Gandhinagar, India in 2020. He will be pursuing a Ph.D. in Computer Science at the University of Utah in the United States of America.
His research interest includes using approximate hardware and other emerging technologies for improving the computing stack.
\end{IEEEbiography}

\begin{IEEEbiography}[{\includegraphics[width=1.1in,height=1.25in]{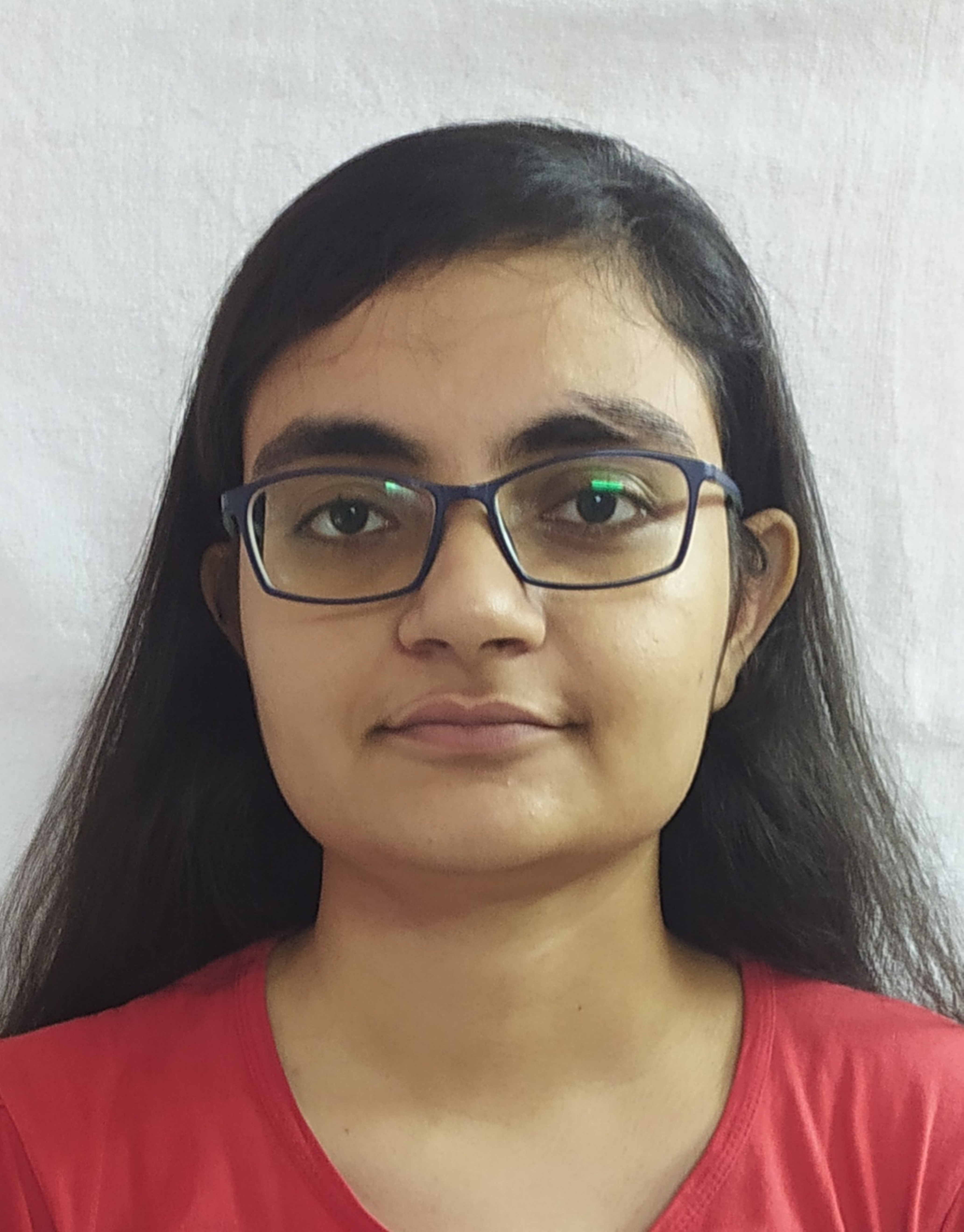}}]
{Riddhi Thakker} received her B.Tech Degree in Information and Communication Technology from Dhirubhai Ambani Institute of Information and Communication Technology, Gandhinagar, Gujarat in 2020. She is currently working as an Applications Engineer in Oracle India. Her research interest includes approximate computing, parallelization using GPU, speech processing and machine learning.
\end{IEEEbiography}

\begin{IEEEbiography}[{\includegraphics[width=1.1in,height=1.25in]{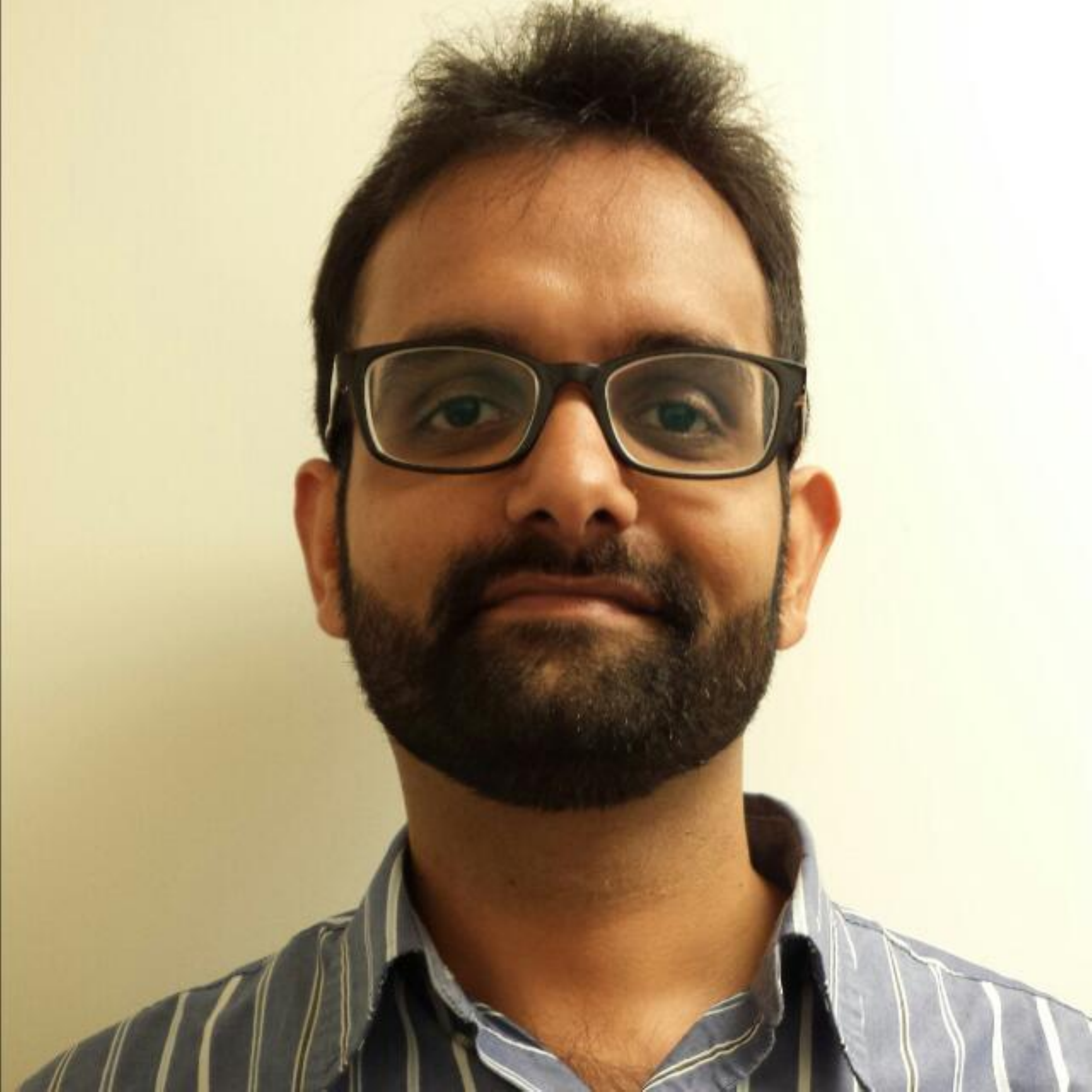}}]
{Manu Awasthi} is an Associate Professor at Ashoka University, India. He received his BTech degree from the Indian Institute of Technology, Varanasi, India, and the PhD degree in computer science from the University of Utah. His research interests are performance evaluation, memory and storage architectures, and characterization of datacenter applications.
\end{IEEEbiography}

\begin{IEEEbiography}[{\includegraphics[width=1.1in,height=1.25in]{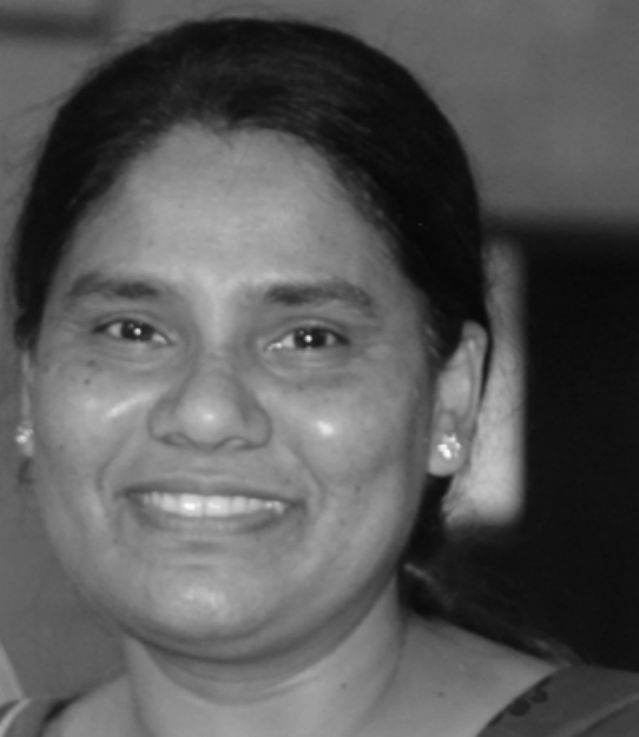}}]{Joycee Mekie} is an Assistant Professor at the Electrical Engineering Department, IIT Gandhinagar. She
received her bachelor’s and master’s degrees in electrical engineering from the M. S. University of Baroda in 1997 and 1999, respectively, and the Ph.D. degree in electrical engineering from IIT Bombay in 2009. Her research interests include approximate computing, circuits for space applications, asynchronous systems. She has served as the reviewer for several journals, including IEEE TCAS I, IEEE TCAS II AND IEEE TCAD.
\end{IEEEbiography}
\end{document}